\documentclass[%
 aip,
 amsmath,amssymb,
 reprint,%
]{revtex4-1}

\usepackage{graphicx}% Include figure files
\usepackage{dcolumn}% Align table columns on decimal point
\usepackage{bm}% bold math
\usepackage[utf8]{inputenc}
\usepackage[T1]{fontenc}
\usepackage{hyperref}

\usepackage{amssymb}
\usepackage{amsmath}
\usepackage{subfigure}

\usepackage{array,multirow}
\usepackage{makecell}
\usepackage{threeparttable}
\usepackage[dvipsnames]{xcolor}

\begin{document}

\preprint{AIP/123-QED}

\title[Point-Cloud Deep Learning of Porous Media for Permeability Prediction]{Point-Cloud Deep Learning of Porous Media for Permeability Prediction}
% Force line breaks with \\

\author{Ali Kashefi}

\email{kashefi@stanford.edu(The author to whom correspondence may be addressed.)}
 \affiliation{Department of Civil and Environmental Engineering, 
Stanford University, Stanford, CA 94305, USA}%Lines break automatically or can be forced with \\

\author{Tapan Mukerji}%
 \email{mukerji@stanford.edu}
\affiliation{ 
Department of Energy Resources Engineering, Stanford University, Stanford, CA 94305, USA
%\\This line break forced with \textbackslash\textbackslash
}%


\begin{abstract}

We propose a novel deep learning framework for predicting permeability of porous media from their digital images. Unlike convolutional neural networks, instead of feeding the whole image volume as inputs to the network, we model the boundary between solid matrix and pore spaces as point clouds and feed them as inputs to a neural network based on the PointNet architecture. This approach overcomes the challenge of memory restriction of graphics processing units and its consequences on the choice of batch size, and convergence. Compared to convolutional neural networks, the proposed deep learning methodology provides freedom to select larger batch sizes, due to reducing significantly the size of network inputs. Specifically, we use the classification branch of PointNet and adjust it for a regression task. As a test case, two and three dimensional synthetic digital rock images are considered. We investigate the effect of different components of our neural network on its performance. We compare our deep learning strategy with a convolutional neural network from various perspectives, specifically for maximum possible batch size. We inspect the generalizability of our network by predicting the permeability of real-world rock samples as well as synthetic digital rocks that are statistically different from the samples used during training. The network predicts the permeability of digital rocks a few thousand times faster than a Lattice Boltzmann solver with a high level of prediction accuracy.

\end{abstract}

\maketitle

%\begin{quotation}
%The ``lead paragraph'' is encapsulated with the \LaTeX\ 
%\verb+quotation+ environment and is formatted as a single paragraph before the first section heading. 
%(The \verb+quotation+ environment reverts to its usual meaning after the first sectioning command.) 
%Note that numbered references are allowed in the lead paragraph.
%
%The lead paragraph will only be found in an article being prepared for the journal \textit{Chaos}.
%\end{quotation}

\section{\label{1}Introduction and motivation}

The importance of study of porous media in a wide range of scientific and industrial fields such as digital rock physics \cite{andra2013digital1,andra2013digital2}, membrane systems \cite{gruber2011computational}, geological carbon storage \cite{blunt2013pore}, and medicine \cite{khanafer2012role} in the past decades has led to a growth in collection of pore-scale image data. Along with pore-scale imaging, there has been a growth in the use of numerical computation to assess physical and transport properties of porous media based on the image data. Such a revolution in the age of data has motivated the use of machine learning schemes as a data-driven strategy to accelerate the computations for understanding physical properties of porous media. Among different machine learning techniques, deep learning has been widely used in various applications for the study of porous media. A few specific applications are: rock image segmentation \cite{karimpouli2019segmentation,da2021deep,niu2020digital}; predicting physical properties and geometrical features such as permeability \cite{graczyk2020predicting,bhatt2002reservoir,hong2020rapid,wu2018seeing,roding2020predicting,tembely2020deep,tian2020permeability,zolotukhin2019machine}, porosity \cite{alqahtani2020machine,rabbani2020deepore,graczyk2020predicting,bordignon2019deep,hebert2020digital}, effective diffusivity \cite{wu2019predicting}, wave propagation velocities \cite{karimpouli2019image}, and fluid flow fields \cite{da2020ml,santos2020poreflow}. It is worth noting that arguments and ideas proposed in this article are general and usable for any desired porous media such as biological tissues and ceramics; however, we restrict ourselves to the applications of rocks in subsurface aquifers and petroleum reservoirs. Specifically, our focus in the present article is on deep learning frameworks for predicting permeability from digital rock images.

Convolutional neural networks (CNNs) have been used extensively to predict the permeability of digital rock images (see e.g., Refs. \onlinecite{wu2018seeing,tembely2020deep,hong2020rapid}). In this setup, CNNs are trained on a set of labeled data to learn a mapping from two or three dimensional digital rock images to rock permeability. Generally speaking, a common challenge in using CNNs is the Graphics Processing Unit (GPU) memory required for training CNNs \cite{li2019novel}. This challenge is magnified in large, deep, and three dimensional CNNs \cite{li2019novel}. Limitation on the memory of available GPU memory might lead to a restriction on the ``batch size'' (see e.g., Ref. \onlinecite{Goodfellow-et-al-2016} for the technical definition of ``batch siz''). Contrarily to the technique of stochastic gradient descent, the mini batch gradient descent method accelerates the training procedure mainly by vectorization. However, the associated batch size affects the performance of deep neural networks \cite{keskar2016large,kandel2020effect,masters2018revisiting,bengio2012practical}. Hence, it is vital to have freedom to choose the optimal batch size. To overcome the above mentioned challenge, we propose a new machine learning architecture, which is based on the deep learning of point cloud data. Next, we explain the key idea of our methodology.

\begin{figure*}[hbt!]
\centering
\includegraphics[width=1.0\linewidth]{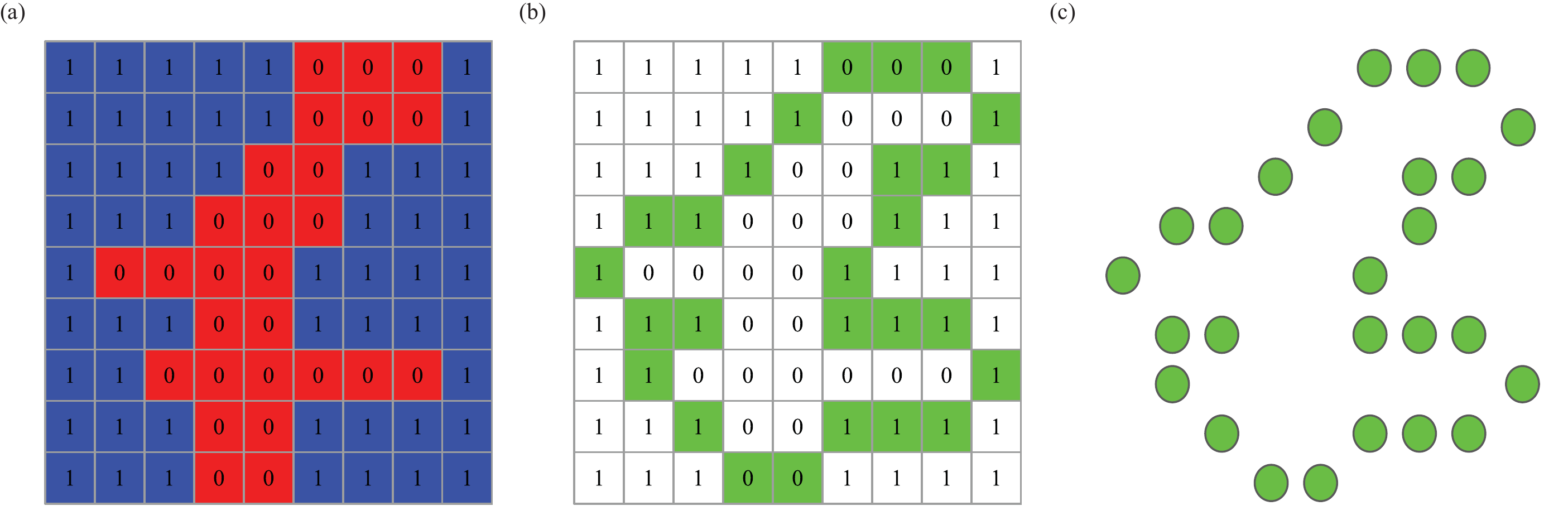}
\caption{
Schematic illustration of the algorithm for constructing point clouds; \textbf{(a)} voxel representation, 0 (red) and 1 (blue) indicate respectively pore and grain spaces \textbf{(b)} pore-grain space boundary identification \textbf{(c)} point could representation, the green boundary is specified by a set of points.}
\label{Fig1}
\end{figure*}

Mathematically, the permeability of a porous medium is a function of the velocity fields in the pore-space of the medium. The solution of the governing equations of fluid flow in porous media (e.g., reservoir rocks) is a function of the geometry of the grain-pore boundary and the boundary conditions. Thus, if the geometry of the grain-pore boundary can be used as an input representation to the neural network, we do not need either the volume of the grain spaces or pore spaces anymore. To reach this goal, for a given porous medium, we only take the grain-pore boundary and represent it as a set of points, constructing a point cloud (see Fig. \ref{Fig1}). Points on the surface (in three dimensional geometries) or on the edge (in two dimensional geometries) of this cloud represent the geometry of the grain-pore boundaries. Representing digital rocks as sparse point clouds instead of full two or three dimensional image pixels or voxels dramatically diminishes the size of memory required to be allocated on GPUs. Additionally, it provides users with more freedom to select the batch size.

Since we represent the boundary of the pore space as a point cloud, a point-cloud-based deep learning framework is required. From a computer science point of view, several architectures are available for this purpose (see e.g., Refs. \onlinecite{wang2019dynamic,qi2017pointnet,thomas2019kpconv}). Among these options, PointNet \cite{qi2017pointnet} has been widely used for deep learning of point cloud data for classification and segmentation of two and three dimensional objects in computer vision and computer graphics (see e.g., Refs. \onlinecite{qi2019deep,qi2018frustum}). \citet{qi2017pointnet} first introduced PointNet in 2017 and the network has quickly become popular for both industrial and academic applications such as object detection \cite{liu2019flownet3d,qi2018frustum}, shape reconstruction \cite{DBLP1:journals/corr/abs-2008-02792}, camera pose estimation \cite{DBLP1:journals/corr/abs-2008-02792}, and physical simulation \cite{kashefi2021point,DBLP2:conf/cvpr/Rempe0WG19}.

To the best of our knowledge, PointNet \cite{qi2017pointnet} has been already used twice for applications outside of the pure computer science areas. First, the performance of PointNet \cite{qi2017pointnet} for predicting the velocity and pressure fields of incompressible flows on irregular geometries has been examined by \citet{kashefi2021point}. They \citet{kashefi2021point} adjusted the PointNet architecture \cite{qi2017pointnet} to predict the flow fields around a cylinder with various shapes for its cross section and obtained an excellent to reasonable level of accuracy. Additionally, they \cite{kashefi2021point} demonstrated the generalizability of their proposed neural network by predicting the velocity and pressure fields on unseen category data such as multiple objects and airfoils (see Figs. 13-19 of Ref. \onlinecite{kashefi2021point}). Second, \citet{defever2019generalized} employed PointNet \cite{qi2017pointnet} to identify local structures in molecular simulations. These successes \cite{defever2019generalized,kashefi2021point} motivate us to utilize the PointNet  \cite{qi2017pointnet} architecture and modify it for our own application. To accomplish this task, we use the classification component of PointNet \cite{qi2017pointnet} and replace its cross-entropy cost function by the mean squared error to establish an end-to-end mapping from a point cloud (as input) to the corresponding permeability (as output) framed as a regression problem. It is important to mention that we utilize PointNet \cite{qi2017pointnet} for the first time for a regression problem. Although, our focus in this article is on permeability prediction in porous media, our approach can be potentially used for any other machine learning problems, where the output of interest is a real number that is a function of the geometry of spatial domains. Further details of our neural network are described in Sect. \ref{231}.

We assess the prediction performance of the network, its sensitivity to different parameters and activation functions, and its computational efficiency in several ways. First, to evaluate prediction performance we report the coefficient of determination as well as the maximum and minimum relative errors of the predicted permeability with reference to the permeability calculated from a numerical solver for a set of two and three dimensional porous medium geometries. Second, to assess the sensitivity to different parameters we discuss the number of points in point clouds as a hyperparameter of the neural network proposed in this article. We evaluate the effect of input and feature transform blocks in PointNet \cite{qi2017pointnet} on the performance of the deep learning framework. Additionally, we explore the influence of different activation functions and different sizes of latent global feature on the accuracy of the predicted permeability, and test the neural network generalizability. Finally, we compute the speed-up factor obtained by the proposed neural network compared to a conventional numerical solver for flow simulation in pore spaces, as well as compare the performance of the point-cloud neural network with a regular CNN.

The rest of this paper is structured as follows. We describe the governing equations of fluid flows in porous media and techniques for permeability computations using numerical solvers in Sect. \ref{21}. Data generation for deep learning is explained in Sect. \ref{22}. We illustrate and compare the architecture of our neural network with a CNN in Sect. \ref{23}. Network training is illustrated in Sect. \ref{24}. An analysis of the network performance for two dimensional geometries is provided in Sect. \ref{31}. Prediction of the permeability in three dimensional porous media is investigated in Sect. \ref{32}. Alternative approaches for permeability prediction and potentials of our neural network in this regard are discussed in Sect. \ref{33}. Section \ref{4} summarizes and concludes the study.

\begin{figure*}[hbt!]
\centering
\includegraphics[width=1.0\linewidth]{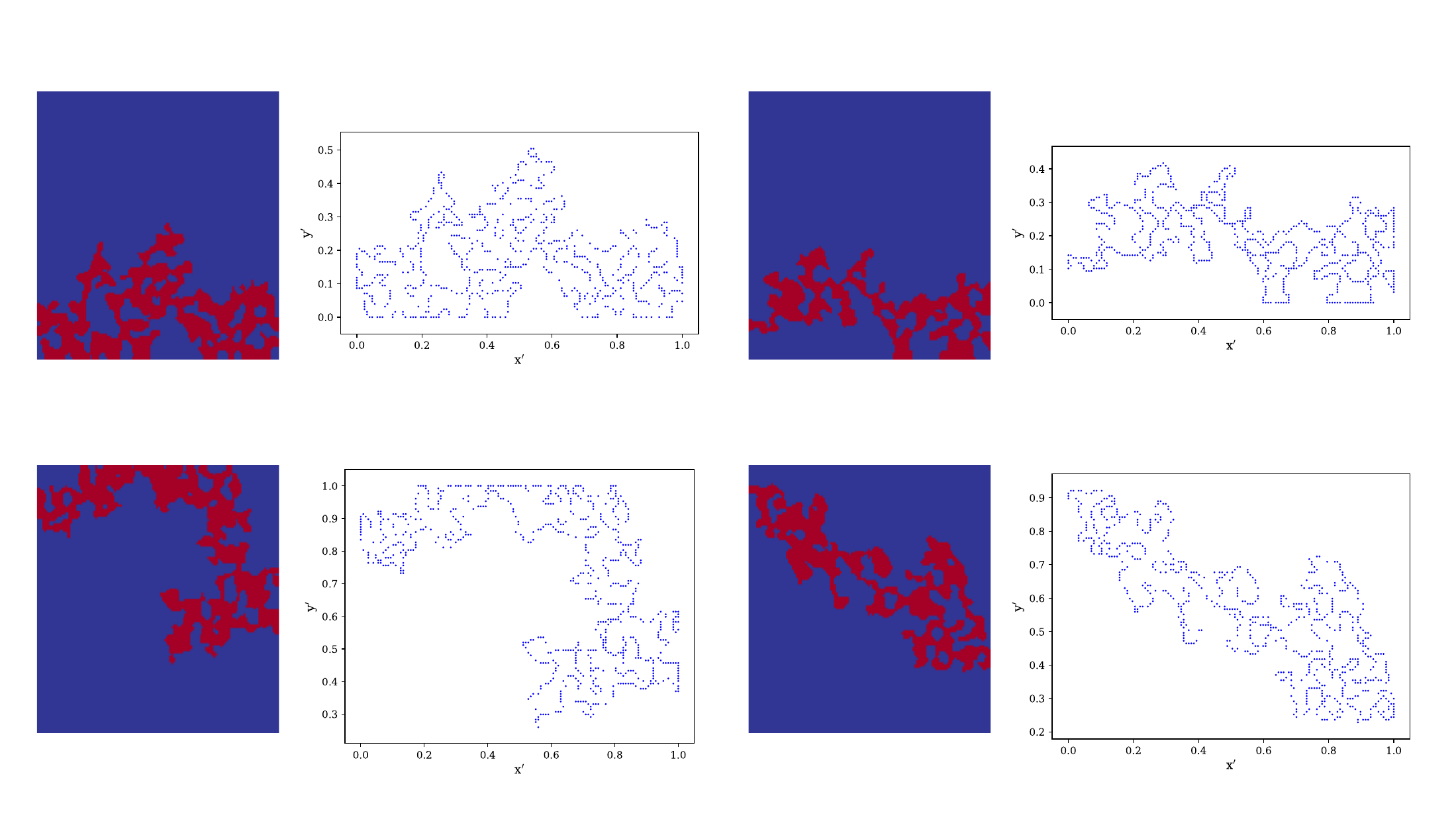}
\caption{Two dimensional digital porous medium images and their corresponding point-cloud representations; digital images and point clouds are used to train CNN and the point-cloud neural network, respectively.}
\label{Fig2}
\end{figure*}

\begin{figure*}[hbt!]
\centering
\includegraphics[width=1.0\linewidth]{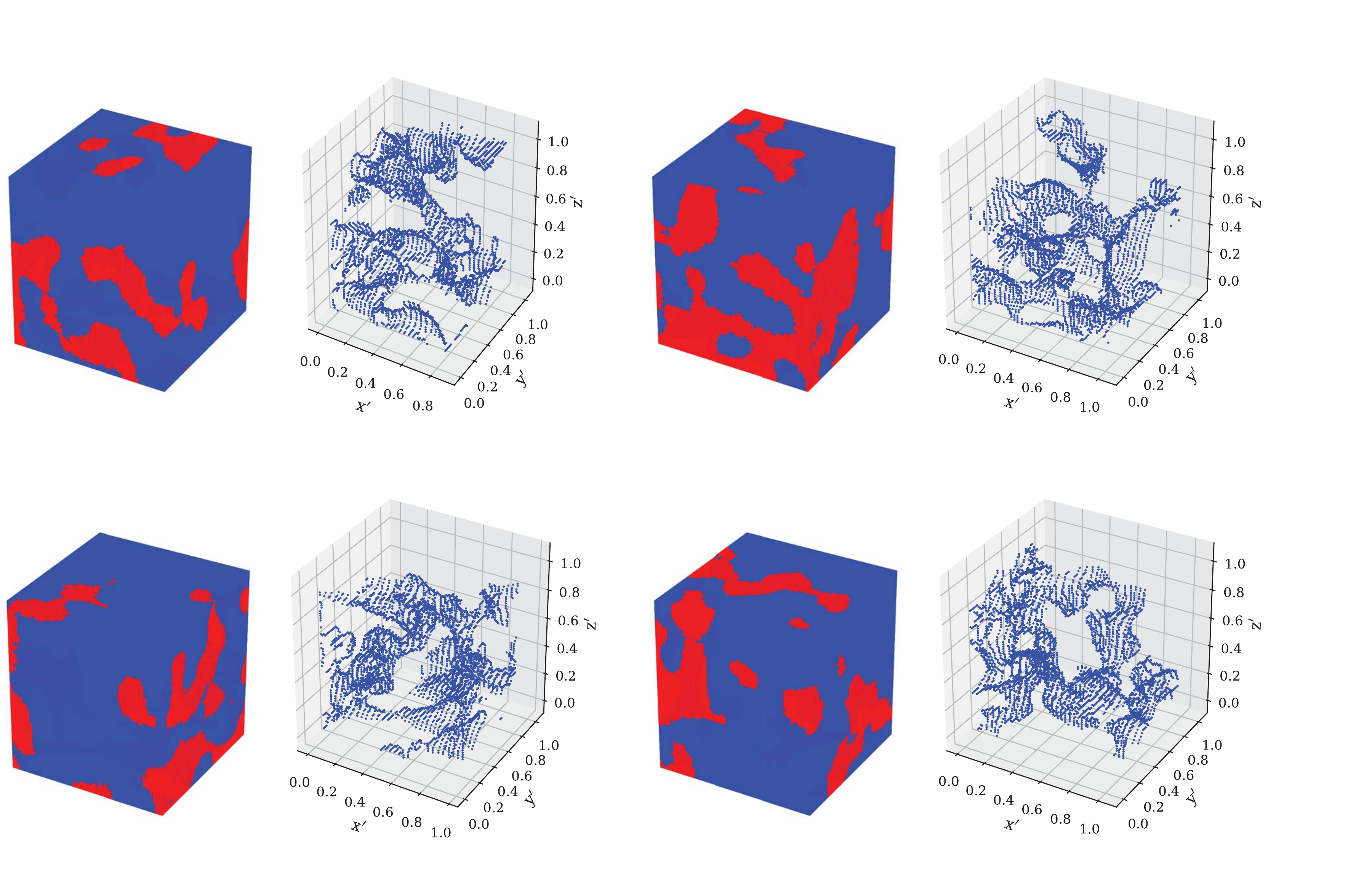}
\caption{Three dimensional digital porous medium images and their corresponding point-cloud representations; digital images and point clouds are used to train CNN and the point-cloud neural network, respectively.}
\label{Fig3}
\end{figure*}

\section{\label{2}Problem formulation and methodology}

\subsection{\label{21}Permeability computation in porous media}

To compute the permeability of a porous medium, first we obtain the velocity field of fluid flow in the pore space of the medium. The continuity and Navier-Stokes equations govern the dynamics of single-phase incompressible flow within the pores of a porous medium and are written as follow:
\begin{eqnarray}
\label{Eq1}
   \frac{\partial \rho}{\partial t} + \nabla \cdot (\rho \textbf{\textit{u}})=0 \textrm{   in   } V, 
\end{eqnarray}
\begin{eqnarray}
\label{Eq2}
\frac{\partial (\rho \textbf{\textit{u}})}{\partial t} + \nabla \cdot (\rho \textbf{\textit{u}} \otimes  \textbf{\textit{u}}) + \nabla p - \mu \Delta \textbf{\textit{u}} = \textbf{\textit{f}} \textrm{   in   } V,
\end{eqnarray}
where $\textbf{\textit{u}}$ and $p$ indicate respectively the velocity vector and absolute pressure in the space of $V$, the fluid-filled pore space. The fluid density and dynamic viscosity are shown by $\rho$ and $\mu$, respectively. The vector of external body force is indicated by $\textbf{\textit{f}}$. We consider the porous medium domains to be squares (in two dimensional spaces) or cubes (in three dimensional spaces) with length $L$ along each principal axis, porosity of $\phi$ and spatial correlation length of $l_c$. A pressure gradient in the $x$ direction ($dp⁄dx$) is applied to stimulate the flow in the medium. No flow boundary condition is enforced at the top and bottom of the medium on the $y$--$z$ planes. Periodic boundary conditions are applied at the inflow and outflow velocity boundaries parallel to the pressure gradient direction. A numerical solver based on the Lattice Boltzmann Method (LBM) is used to obtain the steady state solution to the governing equations. More details of the analysis are discussed in Ref. \onlinecite{keehm2004permeability}. After calculating the flow velocity in the pore space of the porous medium, the permeability in $x$ direction ($k$) is obtained from Darcy’s law \cite{darcy1856fontaines}
\begin{eqnarray}
\label{Eq3}
k=-\frac{\mu \bar{U}}{dp/dx}, 
\end{eqnarray}
where $\bar{U}$ is the mean velocity in the entire porous medium including grain spaces. Note that Eq. \ref{Eq3} is only valid for low Reynolds numbers (see e.g., Ref. \onlinecite{eshghinejadfard2016calculation}).  

\subsection{\label{22}Data generation}

To have a robust control on training data and investigate the effect of different geometrical parameters such as porosity ($\phi$) and spatial correlation length ($l_c$) in porous media, we synthetically generate our data set such that it represents a range of heterogeneity of reservoir rocks. Similar approaches have been taken by \citet{wu2018seeing} and \citet{da2020ml}. To generate a synthetic binary (pore-grain) medium with a targeted porosity ($\phi$) and spatial correlation ($l_c$), a straightforward algorithm of truncated Gaussian simulation \cite{lantuejoul2013geostatistical,xu1993gtsim} is used by truncating spatially correlated Gaussian random fields created by a moving average filter applied to random uncorrelated Gaussian noise. The algorithm is implemented as follows. First, we consider two and three dimensional arrays respectively with the size of $n^2$ and $n^3$. Next, uncorrelated random variables with the standard normal distribution are assigned to the array elements. Afterwards, Gaussian kernels with different kernel sizes are applied as a filter introducing spatial correlation. In the next stage, the numerical values of the arrays are normalized in the range of [0, 1], and thresholded to give binary arrays with desired ranges of porosity (i.e., see Fig. \ref{Fig2} and Fig. \ref{Fig3}). Arrays with no correlated fields are discarded. In this work, we set $L=n\times \delta x$, where $\delta x$ is the size of each pixel side and equal to 0.003 m. Concerning two dimensional porous media, we set $n=128$ and synthetically generate data with three representative spatial correlation lengths (kernel of the Gaussian filter) of 9, 17, and 33 pixels while considering the porosity ($\phi$) in the range of [0.125, 0.25). We use 2600 data samples with a spatial correlation length ($l_c$) of 9 for training, validation, and test purposes, while data with spatial correlation lengths ($l_c$) of 17 and 33 are used for the investigation of neural network generalizability. The mean (and standard deviations) for the porosity and permeability of the training and test set of the two dimensional porous media are as follows: training set porosity, 0.181 (0.03); test set porosity, 0.185 (0.04); training set permeability, 121.62 mD (18.42 mD); test set permeability, 128.52 mD (20.95 mD). Concerning three dimensional porous media, we generate data with $n=64$ and spatial correlation length (kernel of the Gaussian filter) of 17 pixels while the porosity ($\phi$) in the range of [0.125, 0.20) is selected. A total of 2175 samples are generated for use in training, validation, and testing. The mean (and standard deviations) for the porosity and permeability of the training and test set of the three dimensional porous media are as follows: training set porosity, 0.146 (0.04); test set porosity, 0.151 (0.05); training set permeability, 67.12 mD (58.03 mD); test set permeability, 69.83 mD (61.12 mD). We consider real 3D CT-scan image of a rock sample to carry out the generalizability level of our neural network. A set of Python codes and batch files automates the process of generating synthetic data. The LBM solver is run on all of the generated synthetic porous media to get the corresponding permeabilities, thus creating a labeled dataset.

The next step is to define the neural network domain ($V_{NN}$). Indicating the grain-pore boundary by $\partial V$, then mathematically, $V_{NN} \subset \partial V$. In fact, $V_{NN}$ must represent the geometry of the grain-pore boundary. Note that by keeping the physical properties of the fluid (i.e., viscosity and density) and the boundary conditions fixed over all the generated data, the solution of the governing equations is only a function of the geometry of the boundary of the pore space $\partial V$, and consequently $V_{NN}$. $V_{NN}$ contains $N$ points. The challenge is that the number of pixels located on $\partial V$ varies from one data sample to another. Thus, $N$ is a hyperparameter in our deep learning framework. We discuss the effect of the choice of $N$ on our neural network performance in Sect. \ref{31}. Figure \ref{Fig2} and Figure \ref{Fig3} depict respectively digital porous medium images and their resulting point clouds ($V_{NN}$) for two and three dimensions. Note that our deep learning methodology is not limited to constructing $V_{NN}$ from digital images. One may use scattered data obtained on unstructured finite element or finite volume grids to establish $V_{NN}$.

To accelerate the convergence of our neural network training and equalize the contribution of each input component to the training of neural network parameters (e.g., weights and bias), the input and output data are scaled in the range of [0, 1] using the maximum and minimum values of each set of $k$, $x$, $y$, and $z$. We indicate the scaled set by $k^ \prime$, $x^\prime$, $y^\prime$, and $z^\prime$. As an example, $k^\prime$ is computed as follows:
\begin{eqnarray}
\label{Eq4}
k^\prime=\frac{k-\min(k)}{\max(k)-\min(k)}. 
\end{eqnarray}
$x^\prime$, $y^\prime$, and $z^\prime$ are computed similarly. Obviously, $k^\prime$, $x^\prime$, $y^\prime$, and $z^\prime$ are dimensionless.

\subsection{\label{23}Neural network architectures}

\subsubsection{\label{231}Point-cloud neural network}

Our neural network is mainly based on the PointNet \cite{qi2017pointnet} architecture. In this subsection, we briefly describe the point-cloud neural network. One may refer to Ref. \onlinecite{qi2017pointnet} for further explanations. In this subsection, the vectors and matrices of machine learning components are shown by bold letters but not italic. This is to distinguish the machine learning vectors and matrices from the physics-based ones. The two main components are Multilayer Perceptron (MLP) and Fully Connected (FC) layer. An MLP is constructed by several sequential FC layers. We use notation in the form of $(A_1, A_2)$ to show an MLP with two layers, where $A_1$ and $A_2$ are respectively the size of the first and second layer. Notations in the form of $(A_1, A_2, A_3)$ are similarly defined. In the current study, the point-cloud neural network is restricted to MLPs with two and three layers. We parameterize each FC layer by a weight matrix $\mathbf{W}$ and a bias vector $\mathbf{b}$. The size of an FC layer indicates the number of rows in the corresponding matrix $\mathbf{W}$. Mathematically, a recursive function connects the input of $i$th FC layer $\mathbf{a}_i$ to the output of $i-1$th FC layer $\mathbf{a}_{i-1}$ such that
\begin{equation}
\label{Eq5}
\mathbf{a}_i = \sigma(\mathbf{W}_i\mathbf{a}_{i-1} + \mathbf{b}_i),
\end{equation}
where $\sigma$ is a nonlinear activation function. The activation function is applied elementwise to each vector component.

Consider two sets $\mathcal{X}$ and $\mathcal{Y}$, the network inputs and the desired target respectively, with 
$\mathcal{X}=\{\text{x}_i \in \mathbb{R}^d\}_{i = 1}^N$
 and 
$\mathcal{Y}=\{\text{y}_i \in \mathbb{R}\}_{i = 1}^{n_p}$, where $d$ corresponds to the spatial dimension (2 or 3) and $n_p$ is the number of desired physical or geometrical quantities of interest as the targets of the network. When predicting permeability alone, $n_p$ is 1. We wish to design a neural network to map $\mathcal{X}$ to $\mathcal{Y}$ by an operator $f$ such that $\mathcal{Y}=f(\mathcal{X})$. Two fundamental concepts need to be considered in the design. First, the output set $\mathcal{Y}$ is a function of the geometrical features of the input set $\mathcal{X}$. Thus, the operator $f$ must be able to capture the geometrical features. Second, since the input set $\mathcal{X}$ essentially represents an unstructured and unordered point cloud, the operator $f$ must be invariant with respect to the order of input points $\text{x}_i$ of the set $\mathcal{X}$. The PointNet  \cite{qi2017pointnet} architecture provides these two critical features. Hence, we approximate the operator $f$ by a PointNet-based neural network that learns the mapping from $\mathcal{X}$ to $\mathcal{Y}$ through a set of labeled data described in Sect. \ref{22}.

\begin{figure*}[hbt!]
\centering
\includegraphics[width=1.0\linewidth]{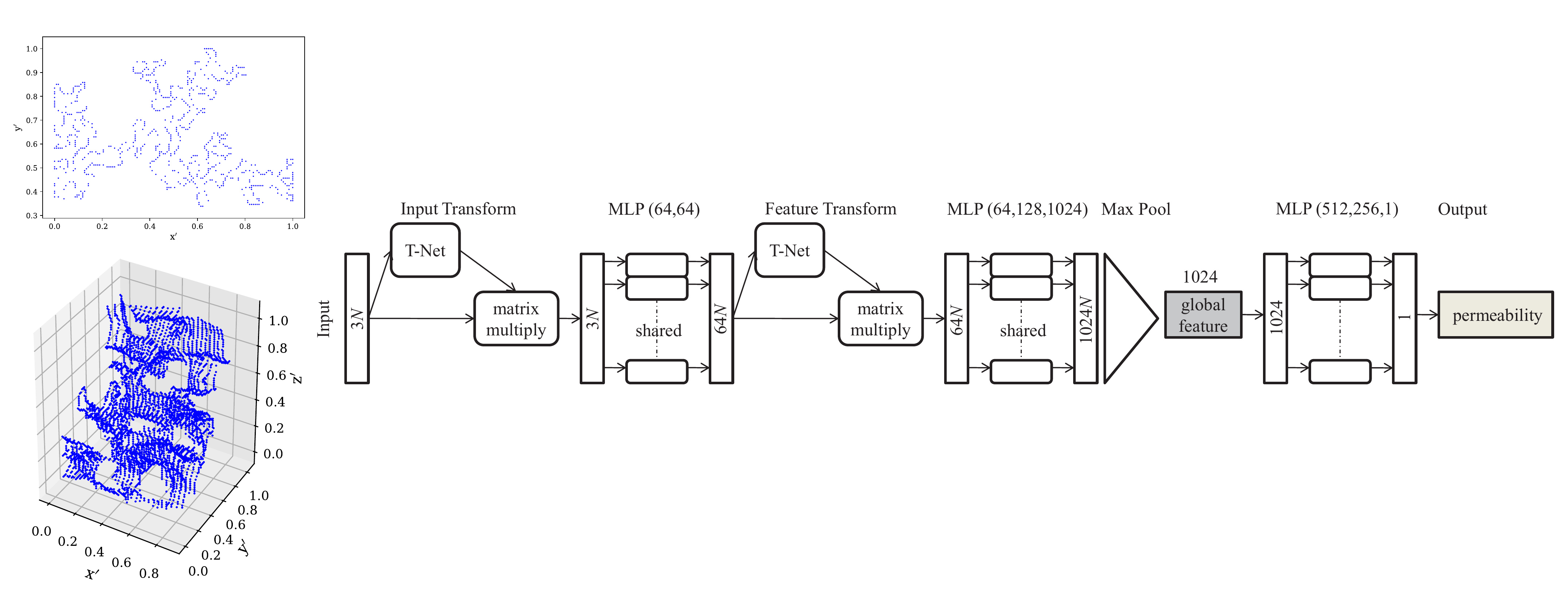}
\caption{Structure of the proposed point-cloud neural network based on PointNet \cite{qi2017pointnet}; the network input is the point cloud representing the boundary line or boundary surface of grain-pore spaces respectively for two and three dimensional porous media.}
\label{Fig4}
\end{figure*}

The structure of the point-cloud neural network is depicted in Fig. \ref{Fig4}. As can be seen in Fig. \ref{Fig4}, the network has two main branches: one before and another after the global feature. The first branch encodes the geometrical feature of the input set $\mathcal{X}$ in a latent global feature with a vector of size 1024. The second branch decodes the latent global feature to predict the permeability. Two Transform Nets (T-Nets) exist in the first branch. The first T-Net transforms the input set $\mathcal{X}$ into an implicit canonical space, while the second T-Net is used for an affine transformation for alignment in the input set $\mathcal{X}$. From a machine learning perspective, T-Nets can be viewed as mini PointNets that consist of an MLP component $(64, 128, 1024)$ followed by a max pooling operator to extract the underlying features. The feature can then be decoded by two MLP components each with two layers $(512, 256)$. One may refer to Ref. \onlinecite{qi2017pointnet} for further descriptions of T-Nets. In addition to T-Nets, two MLP components contribute to the construction of the first branch: the first $(64, 64)$, and the second $(64, 128, 1024)$. Mathematically, PointNet \cite{qi2017pointnet} encodes the geometrical features of the point set such that the latent code is independent of ordering over the set of points. In other words, to aggregate information over the input set $\mathcal{X}$, a permutation invariant function such as maximum, minimum, average, and summation is necessary. PointNet \cite{qi2017pointnet} uses the ``max'' function to handle it. We represent all the mathematical operations carried out on the input set $\mathcal{X}$ just before the max pooling operator by a function $h$. Thus, the latent global feature is established on the input set $\mathcal{X}$ by a function $g$ such that
\begin{equation}
\label{Eq6}
g(\mathcal{X}) \approx \max(h(\text{x}_1), \dots, h(\text{x}_N)).
\end{equation}
As can be observed in Fig. \ref{Fig4}, an MLP component with three layers $(512, 256, 1)$ in the second branch is used to predict the permeability. Note that all the MLP components in the first branch have shared weights, while this is not the case for the MLP component in the second branch. This is another key feature of PointNet \cite{qi2017pointnet} to handle unordered points in the set $\mathcal{X}$; and this is why we use the single function $h$ for all the input points $\text{x}_i$ in Eq. \ref{Eq6}. In fact, it does not matter how the input set $\mathcal{X}$ is constructed for feeding it to our neural network as PointNet \cite{qi2017pointnet} treats all $\text{x}_i$ in a same manner due to the shared weights of MLPs in the first branch. After each FC layer, a batch normalization \cite{ioffe2015batch} operator is used. The activation function used for all the layers is the Rectified Linear Unit (ReLU) defined as
\begin{equation}
\label{Eq7}
\sigma(\gamma) = \max(0, \gamma), 
\end{equation}
except for the last layer, where we employ a sigmoid function expressed as
\begin{equation}
\label{Eq8}
\sigma(\gamma) = \frac{1}{1 +e^{-\gamma}}.
\end{equation}

To close this subsection, we address a few points. First, we set $d=3$ for the permeability prediction in two dimensional pore spaces by assigning zero values to the third axis. Alternatively, one may set $d=2$ for two dimensional porous media and adapt the size of network layers accordingly. Second, we restrict our current study to the prediction of the permeability (i.e., $n_p=1$); however, one may adjust $n_p$ for the prediction of other quantities of interest such as porosity, average pore size, specific surface area, etc (see e.g., Ref. \onlinecite{alqahtani2020machine}).

\subsubsection{\label{232}Convolutional neural networks}

\begin{figure*}[hbt!]
\centering
\includegraphics[width=1.0\linewidth]{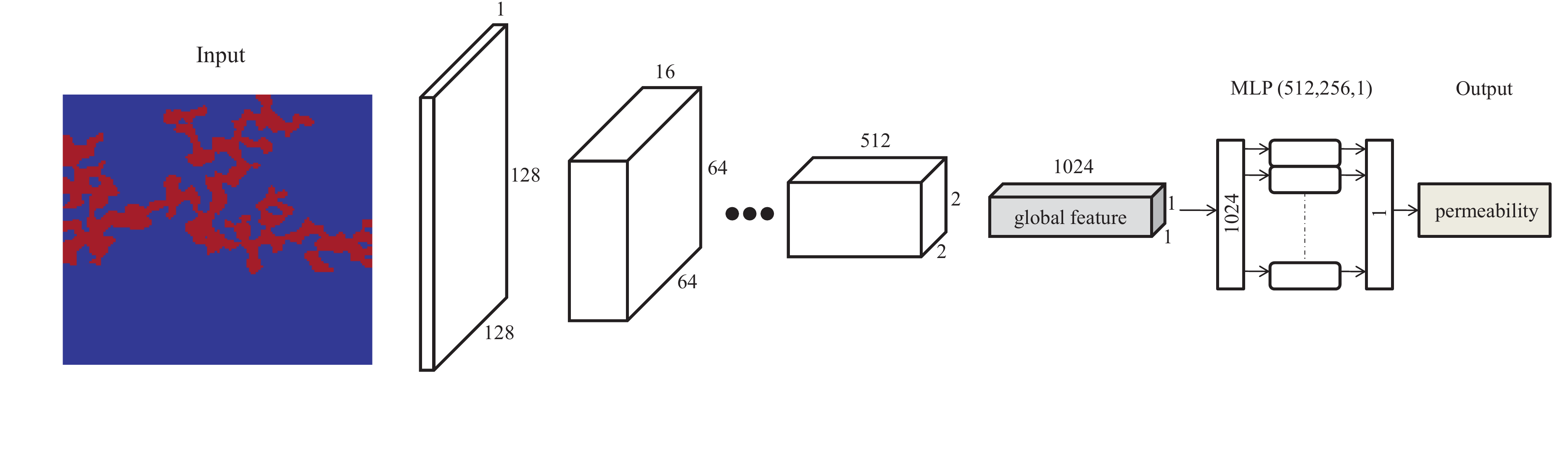}
\caption{Structure of the 2D-CNN used for learning two dimensional porous media}
\label{Fig5}
\end{figure*}

We briefly explain the architecture of the CNNs designed for predicting permeability from two and three dimensional digital rock images. We skip describing the technical details used in this subsection and we encourage potential audiences with interest in use of CNNs for permeability prediction to read Sect. 2.3 of Ref. \onlinecite{hong2020rapid}. Similar to PointNet \cite{qi2017pointnet}, we need an encoder to extract the image features and a decoder, which maps the learned features to the corresponding permeability. We employ the encoder structure of DCGAN \cite{yu2017unsupervised}, which is a highly cited and successful unsupervised generative adversarial network. Accordingly, ReLU activation function is used for all layers and no pooling layer is utilized. The number of filters starts with 16 and doubles at each convolution layer, sequentially. All convolution layers are set with no padding, a stride size of 2, and kernel size of $(2, 2)$ and $(2, 2, 2)$ respectively for the two and three dimensional CNNs, except in the last layer of the three dimensional CNN, which has a kernel size of $(1, 1, 1)$. This is to enforce a latent global feature with the size of 1024 (see further discussions in Sect. \ref{233}). We use the PointNet \cite{qi2017pointnet} decoder for both the two dimensional (2D-CNN) and three dimensional CNN (3D-CNN). As an example, Figure \ref{Fig5} depicts the architecture of 2D-CNN used in this study.

\subsubsection{\label{233}Comparison between PointNet and CNNs}

A fair comparison between PointNet \cite{qi2017pointnet} and a CNN is not straightforward. First, each of them is based on different underlying mathematical and computational theories. Second, PointNet \cite{qi2017pointnet} has a unique structure, whereas we can find many neural networks which are based on the convolution operation (see e.g., Refs. \onlinecite{ajit2020review,khan2020survey}) and fall in the category of CNNs. Moreover, neural networks with the convolution operation are usually combined with other functions such as max-pooling (see e.g., Ref. \onlinecite{nagi2011max}), upsampling (see e.g., Ref. \onlinecite{wang2018understanding}), skip connection (see e.g., Ref. \onlinecite{yamanaka2017fast}), etc. Thus, a variety of CNNs with different performance can be implemented. With these in mind, we have enforced two conditions for designing the CNN introduced in Sect. \ref{232} to make it similar to the PointNet \cite{qi2017pointnet} architecture as much as possible. First, the size of latent global feature of both 2D-CNN and PointNet \cite{qi2017pointnet} is equal to 1024. Second, both networks use the same decoder. This makes it easier to have a consistent comparison.

\subsection{\label{24}Training}

The first step in the training process is to select an appropriate cost function (or loss function). The mean squared error function has been widely used in the area of deep learning of computational mechanics (see e.g., Ref. \onlinecite{sekar2019fast}) as well as in porous media applications (see e.g., Refs. \onlinecite{da2020ml,santos2020poreflow,roding2020predicting,wu2018seeing,hong2020rapid}). In the current study, we utilize this function defined as:
\begin{equation}
\label{Eq9}
\mathcal{C}=\frac{1}{M} \sum_{i=1}^{M} (k_i^{\prime}-\tilde{k}_i^{\prime})^2,
\end{equation}
where $M$ is the number of data in our training set. We label the permeability ($k^ \prime$) obtained by the LBM solver as the “ground truth”, while we denote the predicted permeability by $\tilde{k}^\prime$. After training, we rescale the predicted permeability ($\tilde{k}^\prime$) back into to the physical domain ($k^\prime$) for a post-processing analysis. We use the Adam \cite{kingma2014adam} optimizer with hyperparameters of $\beta_1=0.9$, $\beta_2=0.999$, and $\hat{\epsilon}=10^{-6}$. To understand the mathematical definition of $\beta_1$, $\beta_2$, and $\hat{\epsilon}$, one may refer to Ref. \onlinecite{kingma2014adam}. For two dimensional cases, our generated data are categorized into three sets for training (2300 data), validation (150 data), and test (150 data) through a random selection process. Similarly for three dimensional cases, we have three sets of training (1745 data), validation (215 data), and test (215 data). The validation data set is mainly used to track the convergence rate of the training process and to avoid over-fitting. A systematic procedure through a grid search is undertaken to determine the network hyperparameters. Accordingly, the learning rate of $\alpha=0.07$ for two dimensional cases and $\alpha=0.1$ for three dimensional cases with an exponential decay with the rate of 0.1 provide the optimal choice based on the test cost ($\mathcal{C}$). Using high learning rates ($\alpha$) in neural networks for the permeability prediction has been reported by other researchers as well (e.g., see Fig. 10 of Ref. \onlinecite{tembely2020deep}). We use NVIDIA Tesla V100 graphics card with the memory clock rate of 1.41 GHz and 24 Gigabytes of RAM for the training process. This procedure takes approximately thirty minutes and two hours respectively for two and three dimensional porous media. Note that we only provided the details of training the point-cloud neural network in this subsection. A similar procedure has been taken to obtain the highest possible performance for the CNN discussed in Sect. \ref{232}.

\section{\label{ResultsAndDiscussion}Results and discussion}

\subsection{\label{31}Two dimensional porous media}

The coefficient of determination, namely $R^2$ score, is a commonly used metric to evaluate the performance of predicting the permeability (see e.g., Refs. \onlinecite{hong2020rapid,tembely2020deep,tian2020permeability,wu2018seeing}) and is defined as 
\begin{equation}
\label{Eq10}
R^2=1-\frac{\sum_{i=1}^{P} (k_i-\tilde{k}_i)^2}{\sum_{i=1}^{P} (k_i-\bar{k})^2},
\end{equation}
where $P$ is the number of data in the test set and $\bar{k}$ denotes the mean of the $\{k_i \}_{i=1}^P$ set. We use this metric in the current work.

\begin{figure*}[hbt!]
\centering
\includegraphics[width=1.0\linewidth]{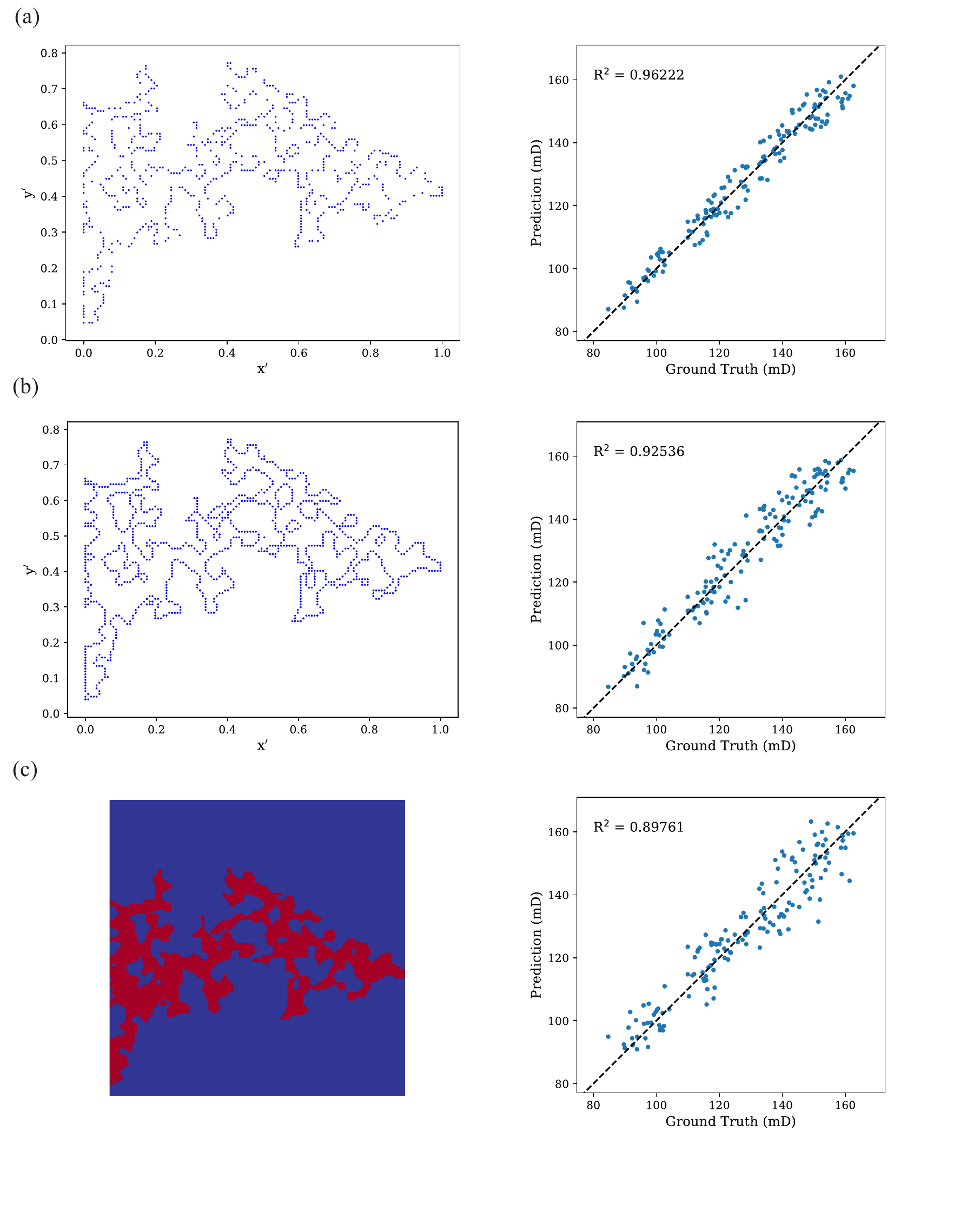}
\caption{Different input representations and their corresponding $R^2$ plots for \textbf{(a)} point-cloud neural network with $N=N_{\text{min}}$, \textbf{(b)} point-cloud neural network with $N=N_{\text{max}}$, and \textbf{(c)} 2D-CNN}
\label{Fig6}
\end{figure*}

The first step in the analysis of our results is to discuss the choice of $N$. As pointed out in Sect. \ref{231}, $N$ is a hyperparameter of our deep learning framework. There is no restriction on $N$ and users of our machine learning platform can tune it to search for the highest achievable performance. For our current digital rock images, the number of points located on the pore-grain boundaries varies between $N_\text{min}=673$ and $N_\text{max}=1698$. To construct $\mathcal{X}$ when $N=N_\text{min}$, we randomly select $N_\text{min}$ points from each point cloud of our data set. Similarly, to establish $\mathcal{X}$ when $N=N_\text{max}$, we add some extra points to point clouds by randomly repeating some of their own points to fill them up to $N_\text{max}$. Additionally, one may select $N$ such that $N_\text{min} < N < N_\text{max}$. Furthermore, there is no restriction on the selections of $N_\text{max}<N$ or $N<N_\text{min}$, although they do not seem reasonable choices, unless one intends to reduce the network size due to a memory limitation by the choice of $N<N_\text{min}$. Figures \ref{Fig6}a and  \ref{Fig6}b respectively depict examples of point cloud illustrations for the choices of $N_\text{min}$ and $N_\text{max}$ and their corresponding $R^2$ score plots. As can be realized from Figs. \ref{Fig6}a and  \ref{Fig6}b, selection of $N=N_\text{min}$ results in a higher $R^2$ score compared to $N=N_\text{max}$ (0.96222 versus 0.92536). From the above described algorithms, we argue that because the $\mathcal{X}$ set contains redundant data in the case of $N=N_\text{max}$, it might cause a deviation in the path of network learning for finding the minimum in the space of the cost function. We also find the $R^2$ scores of 0.90492 and 0.874669 for $N=1000$ and $N=1300$, respectively.

Figure \ref{Fig6}c exhibits the resulting prediction of permeability using the 2D-CNN introduced in Sect. \ref{232}. Compared to our deep learning strategy with $N=N_\text{min}$, a 6.714\% decrease in the $R^2$ score is observed (0.96222 versus 0.89761). A comprehensive comparison between the point-cloud neural network and 2D-CNN is made in Table \ref{Tab1}. Based on the information tabulated in Table  \ref{Tab1}, 2D-CNN experiences higher minimum and maximum relative errors compared to the PointNet based network. Additionally, the size of input vector in CNN increases approximately by a factor of 9, leading to a higher GPU memory requirement. More importantly, the maximum possible batch size on our computational facilities for 2D-CNN is 1024, whereas the point-cloud neural network is able to load all 2300 training data in one epoch. It is conjectured that this is the main reason of a lower performance of 2D-CNN compared to the point-cloud neural network.

\begin{table*}
\caption{\label{Tab1}Comparison between the performance of the point-cloud neural network and 2D-CNN for learning the permeability of two dimensional porous media}
\begin{ruledtabular}
\begin{tabular}{lll}
 & Point-cloud neural network & 2D-CNN \\
\hline
$R^2$ score & 0.96222 & 0.89761  \\
Minimum relative error & 0.002\% & 0.014\%  \\
Maximum relative error & 5.365\% & 13.199\%  \\
Input vector size & 2019 $(N_{\text{min}}\times3)$ & 16384  (128$\times$128 images)\\
Number of trainable parameters & 2,415,763 & 3,459,601\\
\shortstack{Maximum possible batch size\\(increasing by a factor of 2)} & \shortstack{able to load all 2300 training data in one epoch}  & \shortstack{1024}  \\
\end{tabular}
\end{ruledtabular}
\end{table*}

\begin{figure*}[hbt!]
\centering
\includegraphics[width=1.0\linewidth]{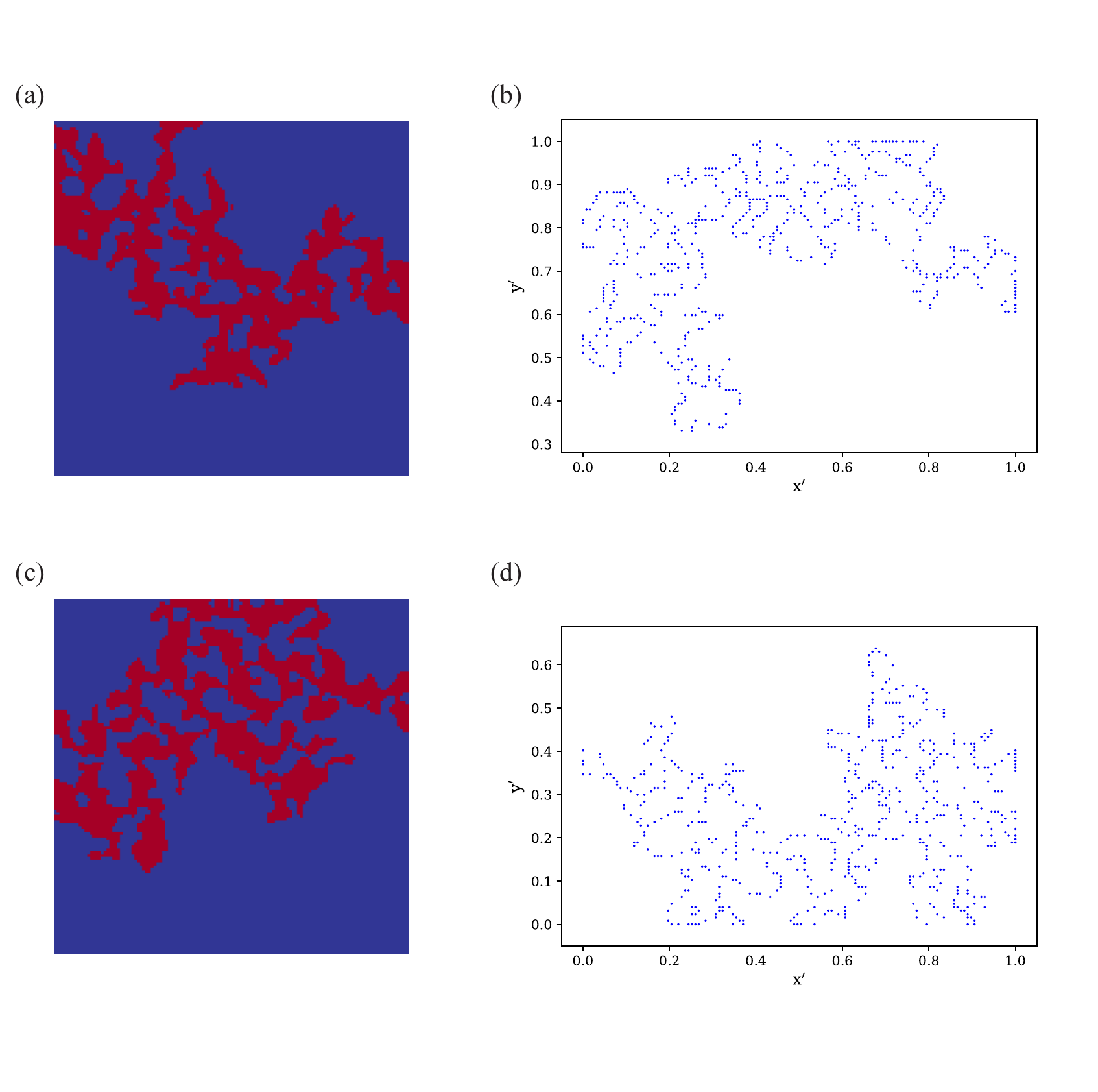}
\caption{Geometries with \textbf{(a)} minimum relative error for 2D-CNN \textbf{(b)} minimum relative error for the point-cloud neural network \textbf{(c)} maximum relative error for 2D-CNN \textbf{(d)} maximum relative error for the point-cloud neural network}
\label{Fig7}
\end{figure*}

Figures \ref{Fig7}a and \ref{Fig7}b illustrate the geometries with the minimum relative errors for the point-cloud neural network and 2D-CNN, while Figures \ref{Fig7}c and \ref{Fig7}d exhibit the geometries with the maximum relative errors for these networks, respectively. As can be inferred from Fig.  \ref{Fig7}, these extremums happen in different geometries for these two networks. It means that each of these two networks has been optimized in two different minima in the high dimensional space of the cost function. Note that we usually do not deal with convex optimization problems in the field of machine learning \cite{jain2017non}. However, because both the maximum and minimum relative errors of the point-cloud neural network are smaller than the corresponding errors of 2D-CNN (see Table \ref{Tab1}), we conclude that the point-cloud neural network is more successful than 2D-CNN to solve the associated optimization problem. Note that as discussed in Sect. \ref{23}, here we report the highest possible performance obtained for each network by a grid search on their hyperparameters. We emphasize on the fact that the goal of this research paper is not to prove that the proposed network can ``definitely'' gain a higher score than any existing CNN-based networks. For instance, one may argue that one can adjust the 2D-CNN proposed in Sect. \ref{232} by making it deeper to reach a higher score compared to our new neural network. Instead, we claim that the PointNet based network with less training efforts and less memory allocations still can compete and outperform CNN-based networks in many cases.

\begin{figure*}[hbt!]
\centering
\includegraphics[width=1.0\linewidth]{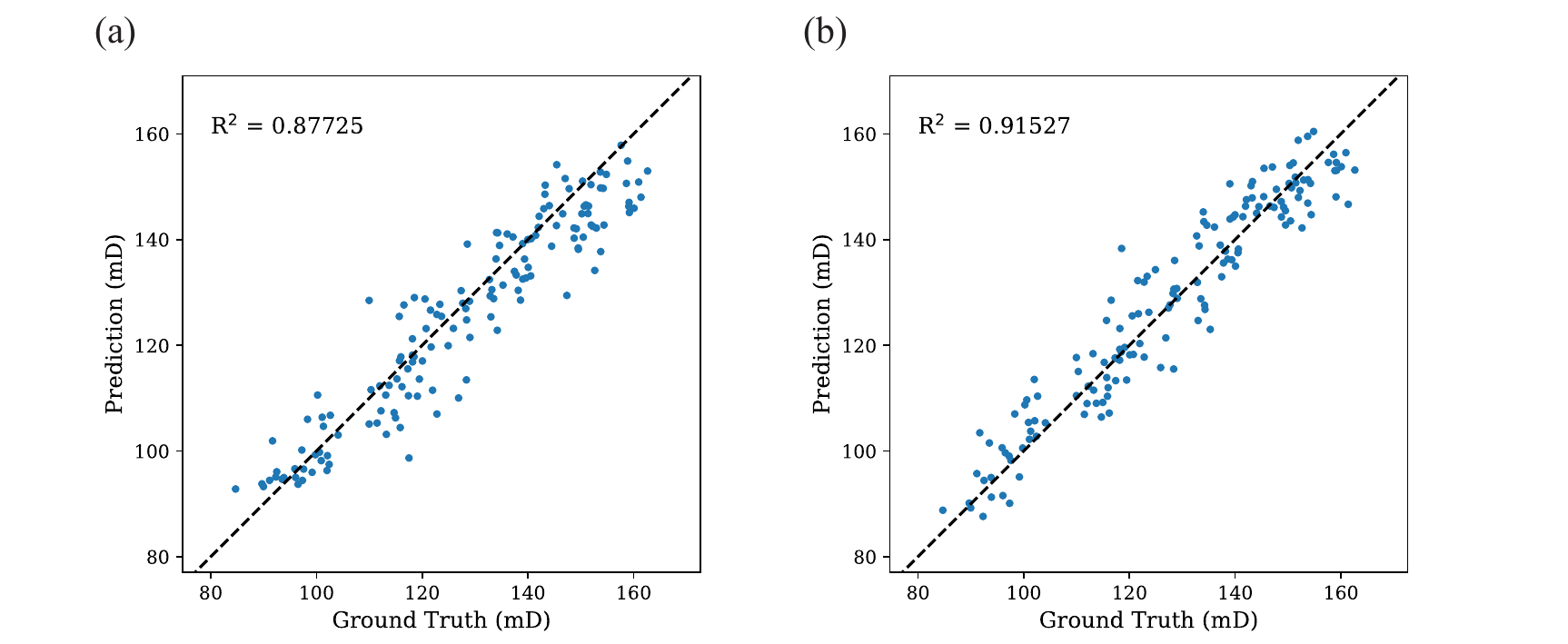}
\caption{$R^2$ scores obtained \textbf{(a)} with the ReLU activation function (see Eq. \ref{Eq7}) in the last layer of the point-cloud neural network; and \textbf{(b)} without using the input and feature transforms in the neural network architecture (see Fig. \ref{Fig4})}
\label{Fig8}
\end{figure*}

As explained in Sect. \ref{22}, we normalize the permeability in the range of [0, 1] for training the network along with the sigmoid activation function (see Eq. \ref{Eq8}) in the last layer of the neural network to cover that range. Our primary motivation to use this approach is that \citet{kashefi2021point} has taken the same procedure for predicting real continuous variables such as velocity and pressure. However, since the permeability is a positive real number another option would be to keep the permeability in the physical domain and use the ReLU activation function (see Eq. \ref{Eq7}) in the last layer. This option has been used by several researchers such as \citet{hong2020rapid} and \citet{tembely2020deep}. We implement the latter option to compare these two strategies. The outcome of using the ReLU function (see Eq. \ref{Eq7}) is illustrated in Fig. \ref{Fig8}a. A comparison between Fig. \ref{Fig8}a and Fig. \ref{Fig6}a indicates a higher $R^2$ score for our current approach (i.e., using the sigmoid activation function). Note that the scatter in Fig. \ref{Fig6} and Fig. \ref{Fig8} is quantified by the $R^2$ scores.

Our next machine learning experiment addresses the effect of input and feature transforms (see Fig. \ref{Fig4}) on the accuracy of predicted permeability. From a computer vision point of view, the input and output transforms have two significant contributions to the shape classification problems. Here, we briefly describe these two contributions at a high level. One may refer to the original PointNet \cite{qi2017pointnet} article for a deeper discussion. First, these two transforms enhance the network performance to identify rotated objects. For instance, a rotated cat still needs to be classified as a cat by PointNet \cite{qi2017pointnet}. Second, using these two transforms the input point clouds are aligned to a canonical space and it leads to a more efficient global feature extraction using the max-pooling operator (see Fig. \ref{Fig4}). Concerning the application considered in this research article, the first contribution mentioned above is not useful. It is mainly because of the fact that we do not rotate our training data for data augmentation purposes. In other words, we are interested in permeability along the $x$--axis (see Eq. \ref{Eq3}) and by a rigid transformation of the digital rock, the corresponding permeability changes in this direction. However, there would be a hope that the second contribution of the transforms to the computer vision application improves our results as well. To answer this question, we remove the input and feature transform blocks from the point-cloud neural network (see Fig. \ref{Fig4}) to investigate its usefulness. Figure \ref{Fig8}b shows the $R^2$ plot as a consequence of this modification. As can be observed in Fig. \ref{Fig8}b, the $R^2$ score is reduced to 0.91527. Hence, we conclude that the existence of these two transforms increases the network ability for permeability prediction, although the network still performs with a relatively high level of accuracy in their absence.

The size of the latent global feature is a critical parameter of PointNet \cite{qi2017pointnet}. \citet{qi2017pointnet} discussed the effect of this parameter on the PointNet \cite{qi2017pointnet} performance for the object classification and segmentation task. \citet{kashefi2021point} also investigated the influence of the global feature size for prediction of the velocity and pressure fields on unstructured grids (see Table II of Ref. \onlinecite{kashefi2021point}). It is important to mention that the original PointNet \cite{qi2017pointnet} is designed with the global feature size of 1024 (see Fig. 2 of Ref. \onlinecite{qi2017pointnet}). Table \ref{Tab2} collects the $R^2$ scores of various global feature sizes for the point-cloud neural network. According to Table \ref{Tab2}, the highest performance is obtained for the size of 1024. Similar results are reported by \citet{qi2017pointnet} and \citet{kashefi2021point}. Note that by changing the global feature size, an adjustment in the size of the MLP right after the latent global feature is necessary to maintain the global feature as the main information bottleneck of the network structure.

\begin{table*}
\caption{\label{Tab2} $R^2$ score as well as minimum and maximum relative errors for two dimensional porous media for different sizes of the global feature with the choice of $N=N_{\text{min}}$; the FC size shows the size of different layers of the fully connected layer right after the global feature in the network (see Fig. \ref{Fig4})}
\begin{ruledtabular}
\begin{tabular}{llllll}
 Global feature size & 128 & 256 & 512 & 1024 & 2048 \\
\hline
FC size & (128, 128, 1) & (256, 128, 1) & (512, 256, 1) & (512, 256, 1) & (512, 256, 1) \\
$R^2$ score & 0.89799 & 0.91651 & 0.91541 & 0.96222 & 0.91845 \\
Minimum relative error & 0.030\% & 0.032\% & 0.004\% & 0.002\% & 0.038\% \\
Maximum relative error & 19.688\% & 13.276\% & 12.294\% &  5.365\% & 13.592\% \\
\end{tabular}
\end{ruledtabular}
\end{table*}

We test the generalizability of the point-cloud neural network by prediction of the permeability of synthetic digital rock images (150 data) with the spatial correlation length of $l_c=17$ and $l_c=33$, while the network has only seen rock images with the correlation length of $l_c=9$ in the training process. Note that from a computer science point of view, the generalizability of a neural network should be examined on unseen data from unseen categories. Thus, measuring the network performance on the test set cannot be interpreted as an indication of network generalizability, because the test set contains unseen data but from seen categories. Table \ref{Tab3} demonstrates the outcome of this test. For $l_c=17$, we only obtain a reasonable level of accuracy with the $R^2$ score of 0.67302. However, for $l_c=33$ the $R^2$ score takes a negative value with a maximum relative error of approximately 30\%. The negative $R^2$ score indicates models that have worse predictions than a baseline based on just the mean value. A similar investigation is conducted for 2D-CNN and a similar trend is experienced. However, a great reduction in the $R^2$ score is observed according to Table \ref{Tab3}. Compared to the point-cloud neural network performance, 2D-CNN experiences higher maximum relative errors and lower minimum relative errors based on the data tabulated in Table \ref{Tab3}. This observation demonstrates that 2D-CNN has relatively high bias on some cases (those predicted by low relative errors) and relatively high variance on other cases (those predicted by high relative errors). From this observation, we can also conclude that 2D-CNN is less generalizable in comparison with the point-cloud neural network. Similar results have been reported by \citet{hong2020rapid}. Although their network \cite{hong2020rapid} trained on Coconino Sandstone achieved the $R^2$ score of 0.872 for the test set of permeability in the $x$--direction, they could only obtain the $R^2$ score of 0.6623 for predicting the same quantity for Bentheim Sandstone.

%However, a greater reduction in the $R^2$ score and maximum relative error for 2D-CNN is observed according to Table \ref{Tab3}.

\begin{table*}
\caption{\label{Tab3} Comparison between the generalizability of the point-cloud neural network and 2D-CNN, where they have never seen samples of porous media with spatial correlation lengths of $l_c=17$ and $l_c=33$ during the training procedure}
\begin{ruledtabular}
\begin{tabular}{lllll}
 & Point-cloud neural network & & 2D-CNN &
\\
\hline
 & $l_c=17$ & $l_c=33$ & $l_c=17$ & $l_c=33$
\\
\hline
$R^2$ score & 0.67302 & -0.91750 & 0.35859 & -1.23118 \\
Minimum relative error & 0.439\% & 0.393\% & 0.048\% & 0.062\% \\
Maximum relative error & 13.178\% & 29.164\% & 66.971\% & 86.507\% \\
\end{tabular}
\end{ruledtabular}
\end{table*}

The next topic to discuss is the speedup factors achieved by our deep learning configuration. The point-cloud neural network estimates the permeability of the test set (150 data) in approximately 6 seconds on the GPU machine available in our computational resources. Computing the permeability of these 150 data using the LBM code written in C\texttt{++} programming language takes on average 1350 seconds (approximately 23 minutes) on a single Intel(R) Core processor with the clock rate of 2.30 GHz. Consequently, the averaged achieved speedup factor is equal to 225 compared to the numerical simulation. Note that the factors reported here are not absolute and strongly depend on the efficiency of the LBM solver and the power of GPU and CPU used. It should also be noted that the LBM solver is an in-house research code running on CPU alone. Modern commercial LBM codes taking advantage of GPUs are expected to be much faster than the code used in this work.

\subsection{\label{32}Three dimensional porous media}

\begin{figure*}[hbt!]
\centering
\includegraphics[width=1.0\linewidth]{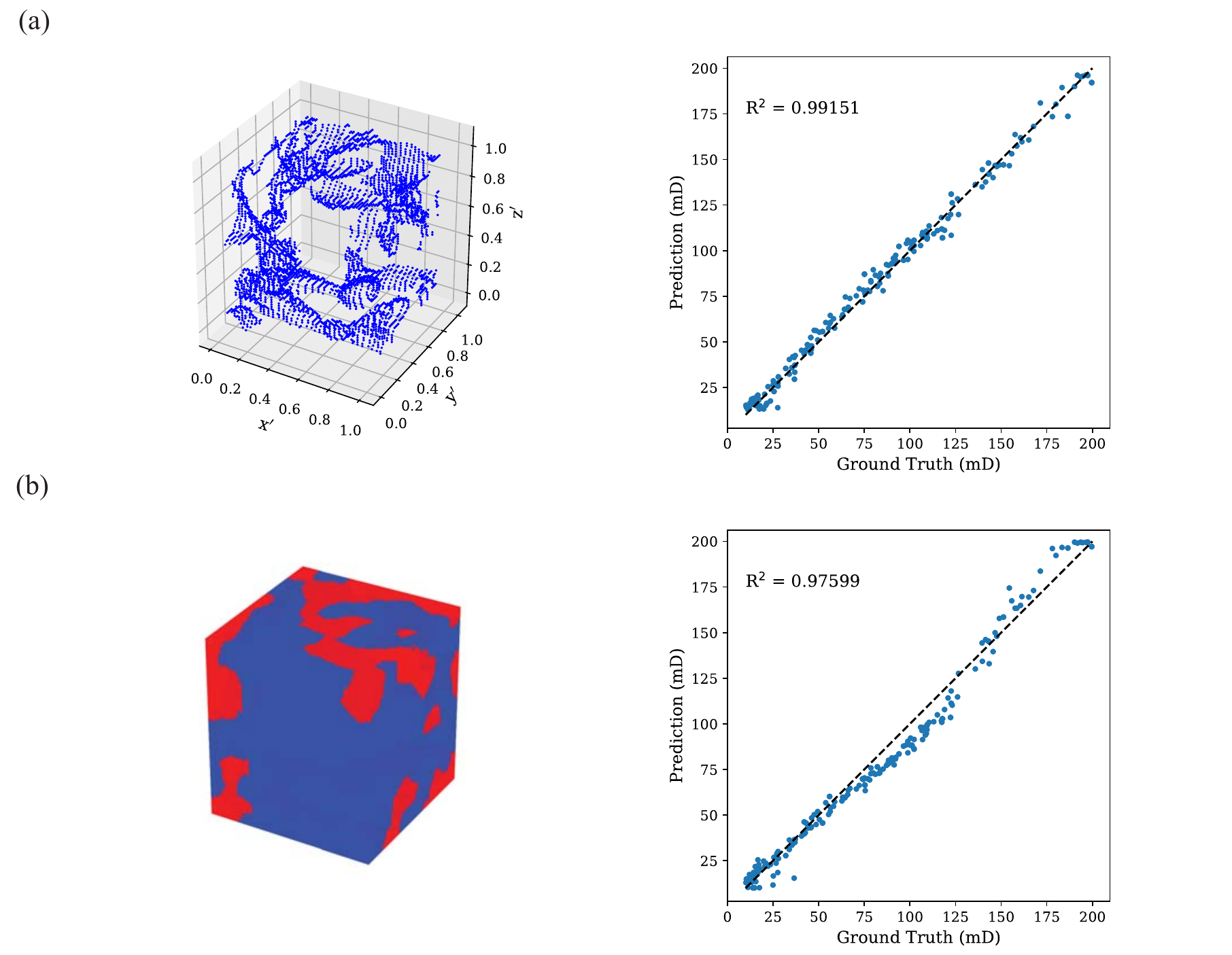}
\caption{Different input representations for three dimensional geometries and their corresponding $R^2$ plots for \textbf{(a)} point-cloud neural network, and \textbf{(b)} 3D-CNN}
\label{Fig9}
\end{figure*}

We construct the three dimensional point clouds (see Fig. \ref{Fig3}) similar to the procedure explained in Sect. \ref{31} with a choice of $N_\text{min}=4003$. Figure \ref{Fig9} compares the $R^2$ score obtained by the point-cloud neural network with that achieved by 3D-CNN. Accordingly, the proposed deep learning technology gains 1.565\% higher accuracy based on the metric of the coefficient of determination (0.99151 versus 0.97599). Table \ref{Tab4} compares these two neural network types from other perspectives. As tabulated in Table \ref{Tab4}, both the minimum and maximum relative errors of 3D-CNN are higher than those obtained by the neural network introduced in this study. In the point-cloud neural network, the minimum and maximum relative errors occur for porous media with permeability of 146.658 mD and 10.148 mD, respectively. The number of trainable parameters of 3D-CNN is slightly greater than the corresponding number in our deep learning framework. Note that the number of trainable parameters of the point-cloud neural network is the same for both the two and three dimensional cases (see Table \ref{Tab1} and Table \ref{Tab4}) and in fact is independent of number of input points. This is simply because the MLP components in the first branch of the point-cloud neural network use shared weights as discussed in Sect. \ref{231}. Note that in contrast to the point-cloud neural network, the number of trainable parameters is a function of input size in CNN architectures.

\begin{table*}
\caption{\label{Tab4}Comparison between the performance of the point-cloud neural network and 3D-CNN for learning the permeability of three dimensional porous media}
\begin{ruledtabular}
\begin{tabular}{lll}
 & Point-cloud neural network & 3D-CNN
\\
\hline
$R^2$ score & 0.99151 & 0.97599  \\
Minimum relative error & 0.030\% & 0.279\%  \\
Maximum relative error & 50.487\% & 57.745\%  \\
Input vector size & 12009 $(N_{min}\times3)$ & 262144 (64$\times$64$\times$64 images)\\
Number of trainable parameters & 2,415,763 & 2,585,169\\
Maximum possible batch size (increasing by a factor of 2) & 2048 & 512  \\
\end{tabular}
\end{ruledtabular}
\end{table*}

\begin{table*}
\caption{\label{Tab5} Effect of batch size on the performance of the point-cloud neural network for predicting the permeability of three dimensional porous media}
\begin{ruledtabular}
\begin{tabular}{llllllllll}
 Batch size & 8 & 16 & 32 & 64 & 128 & 256 & 512 & 1024 & 2048 \\
\hline
$R^2$ score & 0.76297 & 0.70758 & 0.72489 & 0.93464 & 0.98422 & 0.98269 & 0.98466 & 0.99151 & 0.99036\\
Minimum relative error (\%) & 0.013 & 0.032 & 0.060 & 0.224 & 0.019 & 0.058 & 0.066 & 0.030 & 0.031 \\
Maximum relative error (\%) & 264.369 & 578.575 & 104.570 & 60.173 & 64.171 & 43.655 & 75.154 & 50.487 & 40.863 \\
\end{tabular}
\end{ruledtabular}
\end{table*}

\begin{table*}
\caption{\label{Tab6} Effect of batch size on the performance of 3D-CNN for predicting the permeability of three dimensional porous media; the cross symbol ($\times$) indicates that training is impossible due to limitation of GPU memory.}
\begin{ruledtabular}
\begin{tabular}{llllllllll}
 Batch size & 8 & 16 & 32 & 64 & 128 & 256 & 512 & 1024 & 2048 \\
\hline
$R^2$ score & 0.73434 & 0.68299 & 0.92704 & 0.93244 & 0.95803 & 0.96823 & 0.97599 &  $\times$ &  $\times$ \\
Minimum relative error (\%) & 0.096 & 0.300 & 0.079 & 0.320 & 0.007 & 0.097 & 0.279 & $\times$ & $\times$ \\
Maximum relative error (\%) & 116.177 & 80.346 & 80.026 & 73.766 & 77.470 & 79.800 & 57.745 &  $\times$  &  $\times$  \\
\end{tabular}
\end{ruledtabular}
\end{table*}

Based on the information provided in Table \ref{Tab4}, the input of the point-cloud neural network is a vector of size of 12009 ($N_\text{min}\times3$), while this parameter is approximately 22 times greater for 3D-CNN and is equal to 262144 for 64 by 64 by 64 images. This fact leads to the condition that maximum possible usable batch size of the point-cloud neural network is 2048, whereas it is equal to 512 for 3D-CNN. We further investigate the effect of batch size on the performance of the point-cloud neural network and 3D-CNN as reported respectively in Table \ref{Tab5} and Table \ref{Tab6}. As can be realized from Table \ref{Tab5}, the maximum $R^2$ score is obtained with the batch size of 1024 for the point-cloud neural network, while we are not able to run 3D-CNN with the batch size of 1024 and 2048 due to the lack of sufficient GPU memory. According to the data collected in Table \ref{Tab6}, the batch size of 512 results in the maximum $R^2$ score of 3D-CNN, while the performance of 3D-CNN for the batch size of 1024 and 2048 is unknown to us. This is exactly the issue of 3D-CNN addressed in Sect. \ref{1}. In fact, it would be completely possible that 3D-CNN with the batch size of 1024 or 2048 provides a higher $R^2$ score compared to when it is trained with the batch size of 512; however, memory restriction on GPU prevent us trying such experiments. Contrarily, such experiments are doable on the point-cloud neural network and eventually we obtain higher accuracy with the batch size of 1024 compared to 512. We observe how our new methodology overcomes the GPU memory limitation and ends in a higher level of prediction accuracy compared to the CNN methodology.

Concerning the achieved speedup factor, predicting the permeability of the test set (215 data) approximately takes 9 seconds by the point-cloud neural network, while the C\texttt{++} LBM solver computes the permeability of this set in 38700 seconds (approximately 11 hours). Hence, the point-cloud neural network, once trained, accelerates the permeability computations on average by a factor of 4300. Again as mentioned earlier, the LBM solver is an in-house research code running on CPU alone. Modern commercial LBM codes taking advantage of GPUs are expected to be much faster than the code used in this work.

\begin{figure}[hbt!]
\centering
\includegraphics[width=0.7\linewidth]{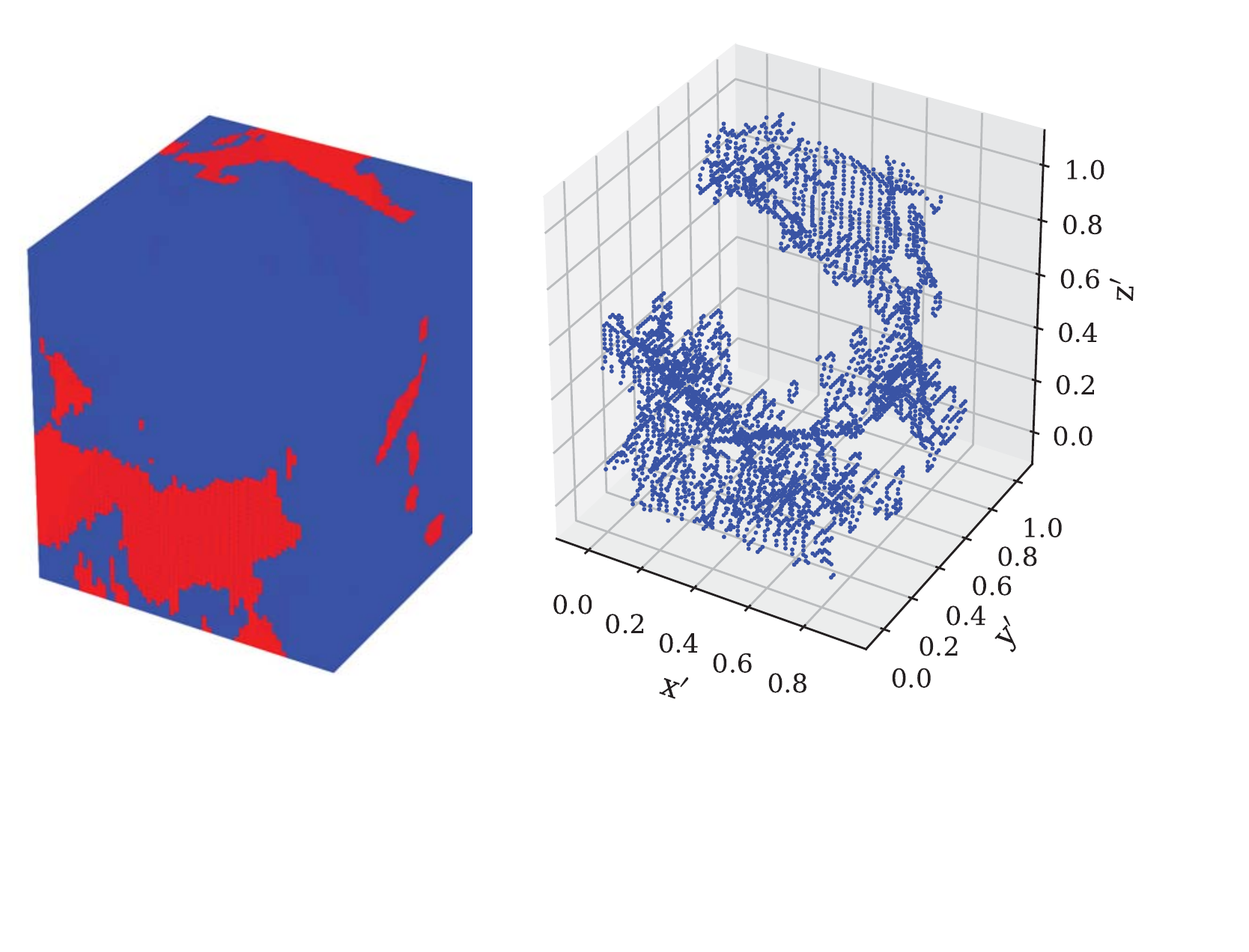}
\caption{Voxel and point cloud representations of one of Berea sandstone samples as a natural porous medium used for exploring the generalizability of our deep learning framework}
\label{Fig10}
\end{figure}

To perform the generalizability of the point-cloud neural network for three dimensional porous media, we inspect the performance of the network on predicting the permeability of Berea sandstone samples (see e.g., Ref. \onlinecite{andra2013digital2}) as natural porous media, while the point-cloud neural network are solely trained on the synthetic data. Figure \ref{Fig10} exhibits the voxel and point cloud representations of one of these samples. The point-cloud neural network obtains $R^2$ score of 0.70437 with respectively the minimum and maximum relative errors of 2.735\% and 35.578\% over 8 samples. We observe that only a reasonable accuracy level is gained because of two main reasons. First, although the permeability of the natural samples is in the range of training data, they have different spatial structure than data during the training procedure. Second, the number of points ($N$) in the clouds constructing the boundary of pore spaces in the natural samples vary between 4388 and 8696, this is while we set $N=4003$ for the network and it ends in losing even further information about the correct structure of the natural samples and thus decreasing the $R^2$ score specifically for samples with large numbers of $N$ (compared to 4003). It is concluded that to obtain higher $R^2$ scores, the network should be trained on similar natural samples or similar synthetic data from permeability, porosity, and spatial correlation length perspectives. Note that the goal of such an experiment is to test the generalizability of the network, meaning that we quantitatively investigate how a deviation from one of these three features negatively affects the accuracy of prediction; however, a decrease in the $R^2$ score would be expected in advance. As discussed in Sect. \ref{31}, we emphasize that the network must be asked to predict unseen data from unseen categories in a generalizability test.

At the end of this subsection, we address three points. First, our machine learning experiments shows that the contribution of input and feature transforms (see Fig. \ref{Fig4}) to increasing the accuracy of predicting the permeability of three dimensional point clouds is insignificant (less than 0.1\%). Thus, one may optionally remove these two transforms from the neural network to make it faster and lighter. Second, an important feature of the point-cloud neural network is its scalability. Depending on the number of points in the training point clouds, one may make the network smaller or larger. Alternatively, one may investigate the effect of the network size on its performance. For example, the size of network can be reduced by scaling its MLPs by a factor of 0.25, which leads to MPLs with sizes of $(16, 16)$, $(16, 32, 128)$, and $(128, 64, 1)$ respectively from left to right as shown in Fig. \ref{Fig4}. In this case, the number of trainable parameters of the network without input and feature transforms decreases from 809601 to 51873. Third, training the point-cloud neural network on a data set containing porous media with a great range of spatial correlation lengths is challenging mainly because such set in practice leads to point-cloud subsets with a comparatively big difference between $N_{max}$ and $N_{min}$. As discussed in Sect. \ref{31}, $N$ is a hyper-parameter that needs to be tuned in the point-cloud neural network; however, when ``$N_{max}-N_{min}$'' is very large, the training process becomes demanding. In fact, by selecting an $N$ close to $N_{min}$, the corresponding geometries of point sets with $N \gg N_{min}$ are poorly represented. On the other hand, by choosing an $N$ close to $N_{max}$, point sets with $N \ll N_{max}$ contain unreasonably a great number of repeated points, which eventually appear as redundant data to the point-cloud network and lead to decreasing the network performance. Moreover, our machine learning experiments show that a moderate $N$ (e.g., arithmetic or geometric mean of $N_{min}$ and $N_{max}$) is not an ideal option as well. Hence, one of our future study plans is to resolve these types of limitations from the proposed point-cloud deep learning framework. It is conjectured that such improvements could positively affect the generalizability of the network as well.

\subsection{\label{33}Potentials for fluid flow field predictions}

In this article, our main focus is the prediction of permeability directly from the digital rock images. However, an alternative approach for the permeability prediction is to first predict the entire velocity fields in the pore spaces using a machine learning framework and then compute the permeability from the predicted velocity field. This approach has been so far taken by several researchers (see e.g., Refs. \onlinecite{da2020ml,santos2020poreflow}). In this approach, a neural network is used as a replacement of conventional numerical solvers to provide an end-to-end mapping from the geometry of porous media to the fluid field of interest in the pore space. Due to high complexity in geometries of natural porous media, an efficient geometry representation in neural networks is necessary. CNNs as deep learning tools have been so far utilized for predicting the velocity fields in porous media. For instance, \citet{santos2020poreflow} used a three dimensional CNN but only to predict one component of the velocity field (parallel to the direction of the applied pressure gradient). \citet{da2020ml} proposed a CNN based on U-Net \cite{ronneberger2015u} to predict the velocity fields in two and three dimensional porous media.

There are two common approaches to represent the geometry of porous media in case of using CNNs. The first approach is to mask the pixels associated with the grains (see e.g., Ref. \onlinecite{santos2020poreflow}). There are two major shortcomings with this approach. The first one is that for each numerical array (representing the porous media) a huge number of pixels have to be masked, specifically for porous media with low porosity. In fact, a considerable portion of computational capacity of CNNs is wasted by taking this strategy. The second one is that for digital rock images with high resolutions, a huge (and uncommon) memory size on Graphics Processing Units (GPUs) is necessary in order to load the entire domain of interest even with a batch size of one. This issue becomes highlighted specifically for prediction of flow fields in three dimensional porous media. For instance, this shortcoming has been reported by \citet{santos2020poreflow} when it was impossible for them to train their own CNN over the entire simulation domain. From a computer vision point of view, a possible solution would be breaking the entire domain into a few subdomains. Although it might resolve the memory issue, it would bring other concerns and constrains for CNNs training and the velocity field prediction. A full discussion on this issue can be found in Sect. 3.1 of Ref. \onlinecite{santos2020poreflow}. The minor issue of this approach is relevant to programming, from a computer science point of view. An efficient function needs to be written to handle the pixel-masking procedure for a large number of training data, each one with a different pattern of fluid flow channels. In this approach, CNNs learn over training data that the fluid flow fields in pore spaces are a function of the number and pattern of the masked pixels. It is important to mention that this technique has been widely used in the area of fluid mechanics for deep learning flows over bluff bodies or airfoils (see e.g., Ref. \onlinecite{bhatnagar2019prediction}). The second approach is to label the input array using two different digit numbers: one representing the pore spaces and another representing the grain (solid) portions. Similarly, the corresponding pixels of pore spaces in the output array takes the numerical value of the velocity field, whereas the corresponding pixels of grain spaces in the output array takes the same number as input. In comparison, with the first approach, this scheme is easier to program and implement and also consumes less wall time over each epoch of training. On the other hand, the main difficulty with this approach is that CNNs not only have to learn the fluid flow fields in pore spaces, but also have to learn the geometry of grain spaces (see e.g., Ref. \onlinecite{da2020ml}). There are three major shortcomings with this approach. The first and second issues are similar to the first approach: wasting considerable computational resources of CNN frameworks and requiring an unreasonable memory size on GPUs even for batching one sample from data set. The third major shortcoming is that machine learning experiments showed that CNNs have troubles identifying the pore spaces versus the grain spaces, specifically when the geometry of flow cluster becomes highly complicated and the effective porosity of the rock decreases. For instance, CNNs proposed by \citet{da2020ml} predicted the velocity fields (regardless the accuracy of its magnitude) in regions where the flow does not even exist (i.e., in grain spaces). Note that there are a few alternative methods to represent the pore spaces as the input of CNNs, such as using the Euclidean distance transform function (see e.g., Refs.  \onlinecite{da2020ml,santos2020poreflow}), instead of using a single number. Although these alternatives might increase the CNN performance, the fundamental issues addressed here remains unchanged. In summary, both of these two approaches suffer from modeling and involving the grain spaces in a CNN deep learning framework. To resolve these issues, we suggest only taking the pore space of a porous medium and representing it as a set of points that constructs a point cloud. Point clouds represent respectively the surface and volume of pore spaces of two and three dimensional porous media. Consequently, one may use the segmentation component of PointNet \cite{qi2017pointnet} to establish an end-to-end mapping between the spatial coordinates of each point of a cloud and the numerical values of the velocity vector at that point. It is conjectured that PointNet\texttt{++} \cite{P++} and Kpconv \cite{thomas2019kpconv} would have higher performance in comparison with PointNet \cite{qi2017pointnet} because they pay more attention to local features of a given geometry.

Moreover, designing an efficient cost function in such problems is critical (see e.g., Sect. 2.3.2 of Ref. \onlinecite{santos2020poreflow}). Cost functions so far used are mainly based on $L^1$ or $L^2$ norm error of the velocity fields \cite{santos2020poreflow,da2020ml}. To enhance the performance of such neural networks and accuracy of the velocity field prediction, we suggest two strategies. Both of these two strategies are based on adding information of the flow governing equations to neural network cost functions. The first strategy leads to a supervised deep learning approach, while the second one results in an unsupervised (or semi-supervised) methodology. The first approach is to add the residual of the continuity and Navier-Stokes equations (Eqs. \ref{Eq1}--\ref{Eq2}) to the cost function. This approach has been carried out in other research areas (see e.g., Refs. \onlinecite{bhatnagar2019prediction,subramaniam2020turbulence}). The second approach is to use the technology of the Physics Informed Neural Network (PINN) \cite{jagtap2020conservative,mao2020physics,raissi2019physics}. In PINNs, the cost function is defined based on the governing equations of the problem of interest as well as the desired initial and boundary conditions, while there is no need of labeled data for training. One may refer to Refs. \onlinecite{jagtap2020conservative,mao2020physics,raissi2019physics} for a deeper discussion on PINNs. Note that there is no mechanism in the current version of PINNs to capture the variations in the geometry of problem domains. In other words, the parameters (e.g., weights and bias) of PINNs are not a function of the geometry of physical domains. Thus, the combination of PointNet \cite{qi2017pointnet} (or other point-cloud based neural networks) with PINNs has the potential to resolve this issue.

\section{\label{4}Summary}

In this study, we introduced a novel point-cloud based deep learning configuration for permeability predictions of digital porous media. We designed the architecture of this configuration according to the classification branch of PointNet \cite{qi2017pointnet}. Taking the advantages of the point-cloud based deep learning methodology, limitations on GPU memory requirements were relaxed and selecting higher batch sizes compared to CNNs became possible. It was mainly due to dramatically diminishing the size of network inputs by only taking the boundary of solid matrix and pore spaces in a porous medium via point cloud representations, rather than taking its whole volume via voxel representations. Freedom in the choice of batch size provided the chance of exploring a relatively wide range of batch sizes to obtain the highest possible accuracy of the permeability prediction. We concentrated on synthetic digital rocks as test cases. According to the metric of coefficient of determination, our deep learning technique achieved an excellent accuracy for the predicted permeability of both two and three dimensional porous media. Compared to a numerical LBM solver, the point-cloud neural network predicted the permeability of test set a few thousand times faster. Finally, we discussed the generalizability of the point-cloud neural network by examining it over two unseen categories: real-world samples; and synthetic samples but with unseen spatial correlation lengths.

\begin{acknowledgments}
We acknowledge the sponsors of the Stanford Center for Earth Resources Forecasting (SCERF) and support from Prof. Steve Graham, the Dean of the Stanford School of Earth, Energy and Environmental Sciences. The work was funded by Shell-Stanford collaborative project on Digital Rock Physics, and the Army Research Office contract \#W911NF1810008. Some of the computing for this project was performed on the Sherlock cluster. We would like to thank Stanford University and the Stanford Research Computing Center for providing computational resources and support that contributed to these research results. Additionally, we wish to thank the reviewers for their insightful comments.
\end{acknowledgments}

\section*{Data availability}
The data that support the findings of this study are available from the corresponding author upon reasonable request.

\section*{References}
%\nocite{*}
\bibliography{aipsamp}% Produces the bibliography via BibTeX.

\begin{thebibliography}{62}%
\makeatletter
\providecommand \@ifxundefined [1]{%
 \@ifx{#1\undefined}
}%
\providecommand \@ifnum [1]{%
 \ifnum #1\expandafter \@firstoftwo
 \else \expandafter \@secondoftwo
 \fi
}%
\providecommand \@ifx [1]{%
 \ifx #1\expandafter \@firstoftwo
 \else \expandafter \@secondoftwo
 \fi
}%
\providecommand \natexlab [1]{#1}%
\providecommand \enquote  [1]{``#1''}%
\providecommand \bibnamefont  [1]{#1}%
\providecommand \bibfnamefont [1]{#1}%
\providecommand \citenamefont [1]{#1}%
\providecommand \href@noop [0]{\@secondoftwo}%
\providecommand \href [0]{\begingroup \@sanitize@url \@href}%
\providecommand \@href[1]{\@@startlink{#1}\@@href}%
\providecommand \@@href[1]{\endgroup#1\@@endlink}%
\providecommand \@sanitize@url [0]{\catcode `\\12\catcode `\$12\catcode
  `\&12\catcode `\#12\catcode `\^12\catcode `\_12\catcode `\%12\relax}%
\providecommand \@@startlink[1]{}%
\providecommand \@@endlink[0]{}%
\providecommand \url  [0]{\begingroup\@sanitize@url \@url }%
\providecommand \@url [1]{\endgroup\@href {#1}{\urlprefix }}%
\providecommand \urlprefix  [0]{URL }%
\providecommand \Eprint [0]{\href }%
\providecommand \doibase [0]{http://dx.doi.org/}%
\providecommand \selectlanguage [0]{\@gobble}%
\providecommand \bibinfo  [0]{\@secondoftwo}%
\providecommand \bibfield  [0]{\@secondoftwo}%
\providecommand \translation [1]{[#1]}%
\providecommand \BibitemOpen [0]{}%
\providecommand \bibitemStop [0]{}%
\providecommand \bibitemNoStop [0]{.\EOS\space}%
\providecommand \EOS [0]{\spacefactor3000\relax}%
\providecommand \BibitemShut  [1]{\csname bibitem#1\endcsname}%
\let\auto@bib@innerbib\@empty
%</preamble>
\bibitem [{\citenamefont {Andr{\"a}}\ \emph
  {et~al.}(2013{\natexlab{a}})\citenamefont {Andr{\"a}}, \citenamefont
  {Combaret}, \citenamefont {Dvorkin}, \citenamefont {Glatt}, \citenamefont
  {Han}, \citenamefont {Kabel}, \citenamefont {Keehm}, \citenamefont
  {Krzikalla}, \citenamefont {Lee}, \citenamefont {Madonna} \emph
  {et~al.}}]{andra2013digital1}%
  \BibitemOpen
  \bibfield  {author} {\bibinfo {author} {\bibfnamefont {H.}~\bibnamefont
  {Andr{\"a}}}, \bibinfo {author} {\bibfnamefont {N.}~\bibnamefont {Combaret}},
  \bibinfo {author} {\bibfnamefont {J.}~\bibnamefont {Dvorkin}}, \bibinfo
  {author} {\bibfnamefont {E.}~\bibnamefont {Glatt}}, \bibinfo {author}
  {\bibfnamefont {J.}~\bibnamefont {Han}}, \bibinfo {author} {\bibfnamefont
  {M.}~\bibnamefont {Kabel}}, \bibinfo {author} {\bibfnamefont
  {Y.}~\bibnamefont {Keehm}}, \bibinfo {author} {\bibfnamefont
  {F.}~\bibnamefont {Krzikalla}}, \bibinfo {author} {\bibfnamefont
  {M.}~\bibnamefont {Lee}}, \bibinfo {author} {\bibfnamefont {C.}~\bibnamefont
  {Madonna}},  \emph {et~al.},\ }\bibfield  {title} {\enquote {\bibinfo {title}
  {Digital rock physics benchmarks—part i: Imaging and segmentation},}\
  }\href@noop {} {\bibfield  {journal} {\bibinfo  {journal} {Computers \&
  Geosciences}\ }\textbf {\bibinfo {volume} {50}},\ \bibinfo {pages} {25--32}
  (\bibinfo {year} {2013}{\natexlab{a}})}\BibitemShut {NoStop}%
\bibitem [{\citenamefont {Andr{\"a}}\ \emph
  {et~al.}(2013{\natexlab{b}})\citenamefont {Andr{\"a}}, \citenamefont
  {Combaret}, \citenamefont {Dvorkin}, \citenamefont {Glatt}, \citenamefont
  {Han}, \citenamefont {Kabel}, \citenamefont {Keehm}, \citenamefont
  {Krzikalla}, \citenamefont {Lee}, \citenamefont {Madonna} \emph
  {et~al.}}]{andra2013digital2}%
  \BibitemOpen
  \bibfield  {author} {\bibinfo {author} {\bibfnamefont {H.}~\bibnamefont
  {Andr{\"a}}}, \bibinfo {author} {\bibfnamefont {N.}~\bibnamefont {Combaret}},
  \bibinfo {author} {\bibfnamefont {J.}~\bibnamefont {Dvorkin}}, \bibinfo
  {author} {\bibfnamefont {E.}~\bibnamefont {Glatt}}, \bibinfo {author}
  {\bibfnamefont {J.}~\bibnamefont {Han}}, \bibinfo {author} {\bibfnamefont
  {M.}~\bibnamefont {Kabel}}, \bibinfo {author} {\bibfnamefont
  {Y.}~\bibnamefont {Keehm}}, \bibinfo {author} {\bibfnamefont
  {F.}~\bibnamefont {Krzikalla}}, \bibinfo {author} {\bibfnamefont
  {M.}~\bibnamefont {Lee}}, \bibinfo {author} {\bibfnamefont {C.}~\bibnamefont
  {Madonna}},  \emph {et~al.},\ }\bibfield  {title} {\enquote {\bibinfo {title}
  {Digital rock physics benchmarks—part ii: Computing effective
  properties},}\ }\href@noop {} {\bibfield  {journal} {\bibinfo  {journal}
  {Computers \& Geosciences}\ }\textbf {\bibinfo {volume} {50}},\ \bibinfo
  {pages} {33--43} (\bibinfo {year} {2013}{\natexlab{b}})}\BibitemShut
  {NoStop}%
\bibitem [{\citenamefont {Gruber}\ \emph {et~al.}(2011)\citenamefont {Gruber},
  \citenamefont {Johnson}, \citenamefont {Tang}, \citenamefont {Jensen},
  \citenamefont {Yde},\ and\ \citenamefont
  {H{\'e}lix-Nielsen}}]{gruber2011computational}%
  \BibitemOpen
  \bibfield  {author} {\bibinfo {author} {\bibfnamefont {M.}~\bibnamefont
  {Gruber}}, \bibinfo {author} {\bibfnamefont {C.}~\bibnamefont {Johnson}},
  \bibinfo {author} {\bibfnamefont {C.}~\bibnamefont {Tang}}, \bibinfo {author}
  {\bibfnamefont {M.}~\bibnamefont {Jensen}}, \bibinfo {author} {\bibfnamefont
  {L.}~\bibnamefont {Yde}}, \ and\ \bibinfo {author} {\bibfnamefont
  {C.}~\bibnamefont {H{\'e}lix-Nielsen}},\ }\bibfield  {title} {\enquote
  {\bibinfo {title} {Computational fluid dynamics simulations of flow and
  concentration polarization in forward osmosis membrane systems},}\
  }\href@noop {} {\bibfield  {journal} {\bibinfo  {journal} {Journal of
  membrane science}\ }\textbf {\bibinfo {volume} {379}},\ \bibinfo {pages}
  {488--495} (\bibinfo {year} {2011})}\BibitemShut {NoStop}%
\bibitem [{\citenamefont {Blunt}\ \emph {et~al.}(2013)\citenamefont {Blunt},
  \citenamefont {Bijeljic}, \citenamefont {Dong}, \citenamefont {Gharbi},
  \citenamefont {Iglauer}, \citenamefont {Mostaghimi}, \citenamefont
  {Paluszny},\ and\ \citenamefont {Pentland}}]{blunt2013pore}%
  \BibitemOpen
  \bibfield  {author} {\bibinfo {author} {\bibfnamefont {M.~J.}\ \bibnamefont
  {Blunt}}, \bibinfo {author} {\bibfnamefont {B.}~\bibnamefont {Bijeljic}},
  \bibinfo {author} {\bibfnamefont {H.}~\bibnamefont {Dong}}, \bibinfo {author}
  {\bibfnamefont {O.}~\bibnamefont {Gharbi}}, \bibinfo {author} {\bibfnamefont
  {S.}~\bibnamefont {Iglauer}}, \bibinfo {author} {\bibfnamefont
  {P.}~\bibnamefont {Mostaghimi}}, \bibinfo {author} {\bibfnamefont
  {A.}~\bibnamefont {Paluszny}}, \ and\ \bibinfo {author} {\bibfnamefont
  {C.}~\bibnamefont {Pentland}},\ }\bibfield  {title} {\enquote {\bibinfo
  {title} {Pore-scale imaging and modelling},}\ }\href@noop {} {\bibfield
  {journal} {\bibinfo  {journal} {Advances in Water resources}\ }\textbf
  {\bibinfo {volume} {51}},\ \bibinfo {pages} {197--216} (\bibinfo {year}
  {2013})}\BibitemShut {NoStop}%
\bibitem [{\citenamefont {Khanafer}, \citenamefont {Cook},\ and\ \citenamefont
  {Marafie}(2012)}]{khanafer2012role}%
  \BibitemOpen
  \bibfield  {author} {\bibinfo {author} {\bibfnamefont {K.}~\bibnamefont
  {Khanafer}}, \bibinfo {author} {\bibfnamefont {K.}~\bibnamefont {Cook}}, \
  and\ \bibinfo {author} {\bibfnamefont {A.}~\bibnamefont {Marafie}},\
  }\bibfield  {title} {\enquote {\bibinfo {title} {The role of porous media in
  modeling fluid flow within hollow fiber membranes of the total artificial
  lung},}\ }\href@noop {} {\bibfield  {journal} {\bibinfo  {journal} {Journal
  of porous media}\ }\textbf {\bibinfo {volume} {15}} (\bibinfo {year}
  {2012})}\BibitemShut {NoStop}%
\bibitem [{\citenamefont {Karimpouli}\ and\ \citenamefont
  {Tahmasebi}(2019{\natexlab{a}})}]{karimpouli2019segmentation}%
  \BibitemOpen
  \bibfield  {author} {\bibinfo {author} {\bibfnamefont {S.}~\bibnamefont
  {Karimpouli}}\ and\ \bibinfo {author} {\bibfnamefont {P.}~\bibnamefont
  {Tahmasebi}},\ }\bibfield  {title} {\enquote {\bibinfo {title} {Segmentation
  of digital rock images using deep convolutional autoencoder networks},}\
  }\href@noop {} {\bibfield  {journal} {\bibinfo  {journal} {Computers \&
  geosciences}\ }\textbf {\bibinfo {volume} {126}},\ \bibinfo {pages}
  {142--150} (\bibinfo {year} {2019}{\natexlab{a}})}\BibitemShut {NoStop}%
\bibitem [{\citenamefont {Da~Wang}\ \emph {et~al.}(2021)\citenamefont
  {Da~Wang}, \citenamefont {Shabaninejad}, \citenamefont {Armstrong},\ and\
  \citenamefont {Mostaghimi}}]{da2021deep}%
  \BibitemOpen
  \bibfield  {author} {\bibinfo {author} {\bibfnamefont {Y.}~\bibnamefont
  {Da~Wang}}, \bibinfo {author} {\bibfnamefont {M.}~\bibnamefont
  {Shabaninejad}}, \bibinfo {author} {\bibfnamefont {R.~T.}\ \bibnamefont
  {Armstrong}}, \ and\ \bibinfo {author} {\bibfnamefont {P.}~\bibnamefont
  {Mostaghimi}},\ }\bibfield  {title} {\enquote {\bibinfo {title} {Deep neural
  networks for improving physical accuracy of 2d and 3d multi-mineral
  segmentation of rock micro-ct images},}\ }\href@noop {} {\bibfield  {journal}
  {\bibinfo  {journal} {Applied Soft Computing}\ }\textbf {\bibinfo {volume}
  {104}},\ \bibinfo {pages} {107185} (\bibinfo {year} {2021})}\BibitemShut
  {NoStop}%
\bibitem [{\citenamefont {Niu}\ \emph {et~al.}(2020)\citenamefont {Niu},
  \citenamefont {Mostaghimi}, \citenamefont {Shabaninejad}, \citenamefont
  {Swietojanski},\ and\ \citenamefont {Armstrong}}]{niu2020digital}%
  \BibitemOpen
  \bibfield  {author} {\bibinfo {author} {\bibfnamefont {Y.}~\bibnamefont
  {Niu}}, \bibinfo {author} {\bibfnamefont {P.}~\bibnamefont {Mostaghimi}},
  \bibinfo {author} {\bibfnamefont {M.}~\bibnamefont {Shabaninejad}}, \bibinfo
  {author} {\bibfnamefont {P.}~\bibnamefont {Swietojanski}}, \ and\ \bibinfo
  {author} {\bibfnamefont {R.~T.}\ \bibnamefont {Armstrong}},\ }\bibfield
  {title} {\enquote {\bibinfo {title} {Digital rock segmentation for
  petrophysical analysis with reduced user bias using convolutional neural
  networks},}\ }\href@noop {} {\bibfield  {journal} {\bibinfo  {journal} {Water
  Resources Research}\ }\textbf {\bibinfo {volume} {56}},\ \bibinfo {pages}
  {e2019WR026597} (\bibinfo {year} {2020})}\BibitemShut {NoStop}%
\bibitem [{\citenamefont {Graczyk}\ and\ \citenamefont
  {Matyka}(2020)}]{graczyk2020predicting}%
  \BibitemOpen
  \bibfield  {author} {\bibinfo {author} {\bibfnamefont {K.~M.}\ \bibnamefont
  {Graczyk}}\ and\ \bibinfo {author} {\bibfnamefont {M.}~\bibnamefont
  {Matyka}},\ }\bibfield  {title} {\enquote {\bibinfo {title} {Predicting
  porosity, permeability, and tortuosity of porous media from images by deep
  learning},}\ }\href@noop {} {\bibfield  {journal} {\bibinfo  {journal}
  {Scientific Reports}\ }\textbf {\bibinfo {volume} {10}},\ \bibinfo {pages}
  {1--11} (\bibinfo {year} {2020})}\BibitemShut {NoStop}%
\bibitem [{\citenamefont {Bhatt}(2002)}]{bhatt2002reservoir}%
  \BibitemOpen
  \bibfield  {author} {\bibinfo {author} {\bibfnamefont {A.}~\bibnamefont
  {Bhatt}},\ }\emph {\bibinfo {title} {Reservoir Properties from Well Logs
  using neural Networks.}},\ \href@noop {} {Ph.D. thesis},\ \bibinfo  {school}
  {Citeseer} (\bibinfo {year} {2002})\BibitemShut {NoStop}%
\bibitem [{\citenamefont {Hong}\ and\ \citenamefont
  {Liu}(2020)}]{hong2020rapid}%
  \BibitemOpen
  \bibfield  {author} {\bibinfo {author} {\bibfnamefont {J.}~\bibnamefont
  {Hong}}\ and\ \bibinfo {author} {\bibfnamefont {J.}~\bibnamefont {Liu}},\
  }\bibfield  {title} {\enquote {\bibinfo {title} {Rapid estimation of
  permeability from digital rock using 3d convolutional neural network},}\
  }\href@noop {} {\bibfield  {journal} {\bibinfo  {journal} {Computational
  Geosciences}\ }\textbf {\bibinfo {volume} {24}},\ \bibinfo {pages}
  {1523--1539} (\bibinfo {year} {2020})}\BibitemShut {NoStop}%
\bibitem [{\citenamefont {Wu}, \citenamefont {Yin},\ and\ \citenamefont
  {Xiao}(2018)}]{wu2018seeing}%
  \BibitemOpen
  \bibfield  {author} {\bibinfo {author} {\bibfnamefont {J.}~\bibnamefont
  {Wu}}, \bibinfo {author} {\bibfnamefont {X.}~\bibnamefont {Yin}}, \ and\
  \bibinfo {author} {\bibfnamefont {H.}~\bibnamefont {Xiao}},\ }\bibfield
  {title} {\enquote {\bibinfo {title} {Seeing permeability from images: fast
  prediction with convolutional neural networks},}\ }\href@noop {} {\bibfield
  {journal} {\bibinfo  {journal} {Science bulletin}\ }\textbf {\bibinfo
  {volume} {63}},\ \bibinfo {pages} {1215--1222} (\bibinfo {year}
  {2018})}\BibitemShut {NoStop}%
\bibitem [{\citenamefont {R{\"o}ding}, \citenamefont {Ma},\ and\ \citenamefont
  {Torquato}(2020)}]{roding2020predicting}%
  \BibitemOpen
  \bibfield  {author} {\bibinfo {author} {\bibfnamefont {M.}~\bibnamefont
  {R{\"o}ding}}, \bibinfo {author} {\bibfnamefont {Z.}~\bibnamefont {Ma}}, \
  and\ \bibinfo {author} {\bibfnamefont {S.}~\bibnamefont {Torquato}},\
  }\bibfield  {title} {\enquote {\bibinfo {title} {Predicting permeability via
  statistical learning on higher-order microstructural information},}\
  }\href@noop {} {\bibfield  {journal} {\bibinfo  {journal} {Scientific
  reports}\ }\textbf {\bibinfo {volume} {10}},\ \bibinfo {pages} {1--17}
  (\bibinfo {year} {2020})}\BibitemShut {NoStop}%
\bibitem [{\citenamefont {Tembely}, \citenamefont {AlSumaiti},\ and\
  \citenamefont {Alameri}(2020)}]{tembely2020deep}%
  \BibitemOpen
  \bibfield  {author} {\bibinfo {author} {\bibfnamefont {M.}~\bibnamefont
  {Tembely}}, \bibinfo {author} {\bibfnamefont {A.~M.}\ \bibnamefont
  {AlSumaiti}}, \ and\ \bibinfo {author} {\bibfnamefont {W.}~\bibnamefont
  {Alameri}},\ }\bibfield  {title} {\enquote {\bibinfo {title} {A deep learning
  perspective on predicting permeability in porous media from network modeling
  to direct simulation},}\ }\href@noop {} {\bibfield  {journal} {\bibinfo
  {journal} {Computational Geosciences}\ }\textbf {\bibinfo {volume} {24}},\
  \bibinfo {pages} {1541--1556} (\bibinfo {year} {2020})}\BibitemShut {NoStop}%
\bibitem [{\citenamefont {Tian}\ \emph {et~al.}(2020)\citenamefont {Tian},
  \citenamefont {Qi}, \citenamefont {Sun}, \citenamefont {Yaseen},\ and\
  \citenamefont {Pham}}]{tian2020permeability}%
  \BibitemOpen
  \bibfield  {author} {\bibinfo {author} {\bibfnamefont {J.}~\bibnamefont
  {Tian}}, \bibinfo {author} {\bibfnamefont {C.}~\bibnamefont {Qi}}, \bibinfo
  {author} {\bibfnamefont {Y.}~\bibnamefont {Sun}}, \bibinfo {author}
  {\bibfnamefont {Z.~M.}\ \bibnamefont {Yaseen}}, \ and\ \bibinfo {author}
  {\bibfnamefont {B.~T.}\ \bibnamefont {Pham}},\ }\bibfield  {title} {\enquote
  {\bibinfo {title} {Permeability prediction of porous media using a
  combination of computational fluid dynamics and hybrid machine learning
  methods},}\ }\href@noop {} {\bibfield  {journal} {\bibinfo  {journal}
  {Engineering with Computers}\ }\textbf {\bibinfo {volume} {2020}} (\bibinfo
  {year} {2020})}\BibitemShut {NoStop}%
\bibitem [{\citenamefont {Zolotukhin}\ and\ \citenamefont
  {Gayubov}(2019)}]{zolotukhin2019machine}%
  \BibitemOpen
  \bibfield  {author} {\bibinfo {author} {\bibfnamefont {A.}~\bibnamefont
  {Zolotukhin}}\ and\ \bibinfo {author} {\bibfnamefont {A.}~\bibnamefont
  {Gayubov}},\ }\bibfield  {title} {\enquote {\bibinfo {title} {Machine
  learning in reservoir permeability prediction and modelling of fluid flow in
  porous media},}\ }in\ \href@noop {} {\emph {\bibinfo {booktitle} {IOP
  Conference Series: Materials Science and Engineering}}},\ Vol.\ \bibinfo
  {volume} {700}\ (\bibinfo {organization} {IOP Publishing},\ \bibinfo {year}
  {2019})\ p.\ \bibinfo {pages} {012023}\BibitemShut {NoStop}%
\bibitem [{\citenamefont {Alqahtani}\ \emph {et~al.}(2020)\citenamefont
  {Alqahtani}, \citenamefont {Alzubaidi}, \citenamefont {Armstrong},
  \citenamefont {Swietojanski},\ and\ \citenamefont
  {Mostaghimi}}]{alqahtani2020machine}%
  \BibitemOpen
  \bibfield  {author} {\bibinfo {author} {\bibfnamefont {N.}~\bibnamefont
  {Alqahtani}}, \bibinfo {author} {\bibfnamefont {F.}~\bibnamefont
  {Alzubaidi}}, \bibinfo {author} {\bibfnamefont {R.~T.}\ \bibnamefont
  {Armstrong}}, \bibinfo {author} {\bibfnamefont {P.}~\bibnamefont
  {Swietojanski}}, \ and\ \bibinfo {author} {\bibfnamefont {P.}~\bibnamefont
  {Mostaghimi}},\ }\bibfield  {title} {\enquote {\bibinfo {title} {Machine
  learning for predicting properties of porous media from 2d x-ray images},}\
  }\href@noop {} {\bibfield  {journal} {\bibinfo  {journal} {Journal of
  Petroleum Science and Engineering}\ }\textbf {\bibinfo {volume} {184}},\
  \bibinfo {pages} {106514} (\bibinfo {year} {2020})}\BibitemShut {NoStop}%
\bibitem [{\citenamefont {Rabbani}\ \emph {et~al.}(2020)\citenamefont
  {Rabbani}, \citenamefont {Babaei}, \citenamefont {Shams}, \citenamefont
  {Da~Wang},\ and\ \citenamefont {Chung}}]{rabbani2020deepore}%
  \BibitemOpen
  \bibfield  {author} {\bibinfo {author} {\bibfnamefont {A.}~\bibnamefont
  {Rabbani}}, \bibinfo {author} {\bibfnamefont {M.}~\bibnamefont {Babaei}},
  \bibinfo {author} {\bibfnamefont {R.}~\bibnamefont {Shams}}, \bibinfo
  {author} {\bibfnamefont {Y.}~\bibnamefont {Da~Wang}}, \ and\ \bibinfo
  {author} {\bibfnamefont {T.}~\bibnamefont {Chung}},\ }\bibfield  {title}
  {\enquote {\bibinfo {title} {Deepore: a deep learning workflow for rapid and
  comprehensive characterization of porous materials},}\ }\href@noop {}
  {\bibfield  {journal} {\bibinfo  {journal} {Advances in Water Resources}\
  }\textbf {\bibinfo {volume} {146}},\ \bibinfo {pages} {103787} (\bibinfo
  {year} {2020})}\BibitemShut {NoStop}%
\bibitem [{\citenamefont {Bordignon}\ \emph {et~al.}(2019)\citenamefont
  {Bordignon}, \citenamefont {Figueiredo}, \citenamefont {Exterkoetter},
  \citenamefont {Rodrigues},\ and\ \citenamefont
  {Correia}}]{bordignon2019deep}%
  \BibitemOpen
  \bibfield  {author} {\bibinfo {author} {\bibfnamefont {F.}~\bibnamefont
  {Bordignon}}, \bibinfo {author} {\bibfnamefont {L.}~\bibnamefont
  {Figueiredo}}, \bibinfo {author} {\bibfnamefont {R.}~\bibnamefont
  {Exterkoetter}}, \bibinfo {author} {\bibfnamefont {B.~B.}\ \bibnamefont
  {Rodrigues}}, \ and\ \bibinfo {author} {\bibfnamefont {M.}~\bibnamefont
  {Correia}},\ }\bibfield  {title} {\enquote {\bibinfo {title} {Deep learning
  for grain size and porosity distributions estimation on micro-ct images},}\
  }in\ \href@noop {} {\emph {\bibinfo {booktitle} {Proceedings of the 16th
  International Congress of the Brazilian Geophysical Society \& Expogef}}}\
  (\bibinfo {year} {2019})\BibitemShut {NoStop}%
\bibitem [{\citenamefont {H{\'e}bert}\ \emph {et~al.}(2020)\citenamefont
  {H{\'e}bert}, \citenamefont {Porcher}, \citenamefont {Planes}, \citenamefont
  {L{\'e}ger}, \citenamefont {Alperovich}, \citenamefont {Goldluecke},
  \citenamefont {Rodriguez},\ and\ \citenamefont
  {Youssef}}]{hebert2020digital}%
  \BibitemOpen
  \bibfield  {author} {\bibinfo {author} {\bibfnamefont {V.}~\bibnamefont
  {H{\'e}bert}}, \bibinfo {author} {\bibfnamefont {T.}~\bibnamefont {Porcher}},
  \bibinfo {author} {\bibfnamefont {V.}~\bibnamefont {Planes}}, \bibinfo
  {author} {\bibfnamefont {M.}~\bibnamefont {L{\'e}ger}}, \bibinfo {author}
  {\bibfnamefont {A.}~\bibnamefont {Alperovich}}, \bibinfo {author}
  {\bibfnamefont {B.}~\bibnamefont {Goldluecke}}, \bibinfo {author}
  {\bibfnamefont {O.}~\bibnamefont {Rodriguez}}, \ and\ \bibinfo {author}
  {\bibfnamefont {S.}~\bibnamefont {Youssef}},\ }\bibfield  {title} {\enquote
  {\bibinfo {title} {Digital core repository coupled with machine learning as a
  tool to classify and assess petrophysical rock properties},}\ }in\ \href@noop
  {} {\emph {\bibinfo {booktitle} {E3S Web of Conferences}}},\ Vol.\ \bibinfo
  {volume} {146}\ (\bibinfo {organization} {EDP Sciences},\ \bibinfo {year}
  {2020})\ p.\ \bibinfo {pages} {01003}\BibitemShut {NoStop}%
\bibitem [{\citenamefont {Wu}\ \emph {et~al.}(2019)\citenamefont {Wu},
  \citenamefont {Fang}, \citenamefont {Kang}, \citenamefont {Tao},\ and\
  \citenamefont {Qiao}}]{wu2019predicting}%
  \BibitemOpen
  \bibfield  {author} {\bibinfo {author} {\bibfnamefont {H.}~\bibnamefont
  {Wu}}, \bibinfo {author} {\bibfnamefont {W.-Z.}\ \bibnamefont {Fang}},
  \bibinfo {author} {\bibfnamefont {Q.}~\bibnamefont {Kang}}, \bibinfo {author}
  {\bibfnamefont {W.-Q.}\ \bibnamefont {Tao}}, \ and\ \bibinfo {author}
  {\bibfnamefont {R.}~\bibnamefont {Qiao}},\ }\bibfield  {title} {\enquote
  {\bibinfo {title} {Predicting effective diffusivity of porous media from
  images by deep learning},}\ }\href@noop {} {\bibfield  {journal} {\bibinfo
  {journal} {Scientific reports}\ }\textbf {\bibinfo {volume} {9}},\ \bibinfo
  {pages} {1--12} (\bibinfo {year} {2019})}\BibitemShut {NoStop}%
\bibitem [{\citenamefont {Karimpouli}\ and\ \citenamefont
  {Tahmasebi}(2019{\natexlab{b}})}]{karimpouli2019image}%
  \BibitemOpen
  \bibfield  {author} {\bibinfo {author} {\bibfnamefont {S.}~\bibnamefont
  {Karimpouli}}\ and\ \bibinfo {author} {\bibfnamefont {P.}~\bibnamefont
  {Tahmasebi}},\ }\bibfield  {title} {\enquote {\bibinfo {title} {Image-based
  velocity estimation of rock using convolutional neural networks},}\
  }\href@noop {} {\bibfield  {journal} {\bibinfo  {journal} {Neural Networks}\
  }\textbf {\bibinfo {volume} {111}},\ \bibinfo {pages} {89--97} (\bibinfo
  {year} {2019}{\natexlab{b}})}\BibitemShut {NoStop}%
\bibitem [{\citenamefont {Da~Wang}\ \emph {et~al.}(2020)\citenamefont
  {Da~Wang}, \citenamefont {Chung}, \citenamefont {Armstrong},\ and\
  \citenamefont {Mostaghimi}}]{da2020ml}%
  \BibitemOpen
  \bibfield  {author} {\bibinfo {author} {\bibfnamefont {Y.}~\bibnamefont
  {Da~Wang}}, \bibinfo {author} {\bibfnamefont {T.}~\bibnamefont {Chung}},
  \bibinfo {author} {\bibfnamefont {R.~T.}\ \bibnamefont {Armstrong}}, \ and\
  \bibinfo {author} {\bibfnamefont {P.}~\bibnamefont {Mostaghimi}},\ }\bibfield
   {title} {\enquote {\bibinfo {title} {Ml-lbm: Machine learning aided flow
  simulation in porous media},}\ }\href@noop {} {\bibfield  {journal} {\bibinfo
   {journal} {arXiv preprint arXiv:2004.11675}\ } (\bibinfo {year}
  {2020})}\BibitemShut {NoStop}%
\bibitem [{\citenamefont {Santos}\ \emph {et~al.}(2020)\citenamefont {Santos},
  \citenamefont {Xu}, \citenamefont {Jo}, \citenamefont {Landry}, \citenamefont
  {Prodanovi{\'c}},\ and\ \citenamefont {Pyrcz}}]{santos2020poreflow}%
  \BibitemOpen
  \bibfield  {author} {\bibinfo {author} {\bibfnamefont {J.~E.}\ \bibnamefont
  {Santos}}, \bibinfo {author} {\bibfnamefont {D.}~\bibnamefont {Xu}}, \bibinfo
  {author} {\bibfnamefont {H.}~\bibnamefont {Jo}}, \bibinfo {author}
  {\bibfnamefont {C.~J.}\ \bibnamefont {Landry}}, \bibinfo {author}
  {\bibfnamefont {M.}~\bibnamefont {Prodanovi{\'c}}}, \ and\ \bibinfo {author}
  {\bibfnamefont {M.~J.}\ \bibnamefont {Pyrcz}},\ }\bibfield  {title} {\enquote
  {\bibinfo {title} {Poreflow-net: A 3d convolutional neural network to predict
  fluid flow through porous media},}\ }\href@noop {} {\bibfield  {journal}
  {\bibinfo  {journal} {Advances in Water Resources}\ }\textbf {\bibinfo
  {volume} {138}},\ \bibinfo {pages} {103539} (\bibinfo {year}
  {2020})}\BibitemShut {NoStop}%
\bibitem [{\citenamefont {Li}\ \emph {et~al.}(2019)\citenamefont {Li},
  \citenamefont {Shen}, \citenamefont {Dou}, \citenamefont {Ni}, \citenamefont
  {Xu}, \citenamefont {Yang}, \citenamefont {Wang},\ and\ \citenamefont
  {Niu}}]{li2019novel}%
  \BibitemOpen
  \bibfield  {author} {\bibinfo {author} {\bibfnamefont {S.}~\bibnamefont
  {Li}}, \bibinfo {author} {\bibfnamefont {X.}~\bibnamefont {Shen}}, \bibinfo
  {author} {\bibfnamefont {Y.}~\bibnamefont {Dou}}, \bibinfo {author}
  {\bibfnamefont {S.}~\bibnamefont {Ni}}, \bibinfo {author} {\bibfnamefont
  {J.}~\bibnamefont {Xu}}, \bibinfo {author} {\bibfnamefont {K.}~\bibnamefont
  {Yang}}, \bibinfo {author} {\bibfnamefont {Q.}~\bibnamefont {Wang}}, \ and\
  \bibinfo {author} {\bibfnamefont {X.}~\bibnamefont {Niu}},\ }\bibfield
  {title} {\enquote {\bibinfo {title} {A novel memory-scheduling strategy for
  large convolutional neural network on memory-limited devices},}\ }\href@noop
  {} {\bibfield  {journal} {\bibinfo  {journal} {Computational intelligence and
  neuroscience}\ }\textbf {\bibinfo {volume} {2019}} (\bibinfo {year}
  {2019})}\BibitemShut {NoStop}%
\bibitem [{\citenamefont {Goodfellow}, \citenamefont {Bengio},\ and\
  \citenamefont {Courville}(2016)}]{Goodfellow-et-al-2016}%
  \BibitemOpen
  \bibfield  {author} {\bibinfo {author} {\bibfnamefont {I.}~\bibnamefont
  {Goodfellow}}, \bibinfo {author} {\bibfnamefont {Y.}~\bibnamefont {Bengio}},
  \ and\ \bibinfo {author} {\bibfnamefont {A.}~\bibnamefont {Courville}},\
  }\href@noop {} {\emph {\bibinfo {title} {Deep Learning}}}\ (\bibinfo
  {publisher} {MIT Press},\ \bibinfo {year} {2016})\ \bibinfo {note}
  {\url{http://www.deeplearningbook.org}}\BibitemShut {NoStop}%
\bibitem [{\citenamefont {Keskar}\ \emph {et~al.}(2016)\citenamefont {Keskar},
  \citenamefont {Mudigere}, \citenamefont {Nocedal}, \citenamefont
  {Smelyanskiy},\ and\ \citenamefont {Tang}}]{keskar2016large}%
  \BibitemOpen
  \bibfield  {author} {\bibinfo {author} {\bibfnamefont {N.~S.}\ \bibnamefont
  {Keskar}}, \bibinfo {author} {\bibfnamefont {D.}~\bibnamefont {Mudigere}},
  \bibinfo {author} {\bibfnamefont {J.}~\bibnamefont {Nocedal}}, \bibinfo
  {author} {\bibfnamefont {M.}~\bibnamefont {Smelyanskiy}}, \ and\ \bibinfo
  {author} {\bibfnamefont {P.~T.~P.}\ \bibnamefont {Tang}},\ }\bibfield
  {title} {\enquote {\bibinfo {title} {On large-batch training for deep
  learning: Generalization gap and sharp minima},}\ }\href@noop {} {\bibfield
  {journal} {\bibinfo  {journal} {arXiv preprint arXiv:1609.04836}\ } (\bibinfo
  {year} {2016})}\BibitemShut {NoStop}%
\bibitem [{\citenamefont {Kandel}\ and\ \citenamefont
  {Castelli}(2020)}]{kandel2020effect}%
  \BibitemOpen
  \bibfield  {author} {\bibinfo {author} {\bibfnamefont {I.}~\bibnamefont
  {Kandel}}\ and\ \bibinfo {author} {\bibfnamefont {M.}~\bibnamefont
  {Castelli}},\ }\bibfield  {title} {\enquote {\bibinfo {title} {The effect of
  batch size on the generalizability of the convolutional neural networks on a
  histopathology dataset},}\ }\href@noop {} {\bibfield  {journal} {\bibinfo
  {journal} {ICT express}\ }\textbf {\bibinfo {volume} {6}},\ \bibinfo {pages}
  {312--315} (\bibinfo {year} {2020})}\BibitemShut {NoStop}%
\bibitem [{\citenamefont {Masters}\ and\ \citenamefont
  {Luschi}(2018)}]{masters2018revisiting}%
  \BibitemOpen
  \bibfield  {author} {\bibinfo {author} {\bibfnamefont {D.}~\bibnamefont
  {Masters}}\ and\ \bibinfo {author} {\bibfnamefont {C.}~\bibnamefont
  {Luschi}},\ }\bibfield  {title} {\enquote {\bibinfo {title} {Revisiting small
  batch training for deep neural networks},}\ }\href@noop {} {\bibfield
  {journal} {\bibinfo  {journal} {arXiv preprint arXiv:1804.07612}\ } (\bibinfo
  {year} {2018})}\BibitemShut {NoStop}%
\bibitem [{\citenamefont {Bengio}(2012)}]{bengio2012practical}%
  \BibitemOpen
  \bibfield  {author} {\bibinfo {author} {\bibfnamefont {Y.}~\bibnamefont
  {Bengio}},\ }\bibfield  {title} {\enquote {\bibinfo {title} {Practical
  recommendations for gradient-based training of deep architectures},}\ }in\
  \href@noop {} {\emph {\bibinfo {booktitle} {Neural networks: Tricks of the
  trade}}}\ (\bibinfo  {publisher} {Springer},\ \bibinfo {year} {2012})\ pp.\
  \bibinfo {pages} {437--478}\BibitemShut {NoStop}%
\bibitem [{\citenamefont {Wang}\ \emph {et~al.}(2019)\citenamefont {Wang},
  \citenamefont {Sun}, \citenamefont {Liu}, \citenamefont {Sarma},
  \citenamefont {Bronstein},\ and\ \citenamefont {Solomon}}]{wang2019dynamic}%
  \BibitemOpen
  \bibfield  {author} {\bibinfo {author} {\bibfnamefont {Y.}~\bibnamefont
  {Wang}}, \bibinfo {author} {\bibfnamefont {Y.}~\bibnamefont {Sun}}, \bibinfo
  {author} {\bibfnamefont {Z.}~\bibnamefont {Liu}}, \bibinfo {author}
  {\bibfnamefont {S.~E.}\ \bibnamefont {Sarma}}, \bibinfo {author}
  {\bibfnamefont {M.~M.}\ \bibnamefont {Bronstein}}, \ and\ \bibinfo {author}
  {\bibfnamefont {J.~M.}\ \bibnamefont {Solomon}},\ }\bibfield  {title}
  {\enquote {\bibinfo {title} {Dynamic graph cnn for learning on point
  clouds},}\ }\href@noop {} {\bibfield  {journal} {\bibinfo  {journal} {Acm
  Transactions On Graphics (tog)}\ }\textbf {\bibinfo {volume} {38}},\ \bibinfo
  {pages} {1--12} (\bibinfo {year} {2019})}\BibitemShut {NoStop}%
\bibitem [{\citenamefont {Qi}\ \emph {et~al.}(2017{\natexlab{a}})\citenamefont
  {Qi}, \citenamefont {Su}, \citenamefont {Mo},\ and\ \citenamefont
  {Guibas}}]{qi2017pointnet}%
  \BibitemOpen
  \bibfield  {author} {\bibinfo {author} {\bibfnamefont {C.~R.}\ \bibnamefont
  {Qi}}, \bibinfo {author} {\bibfnamefont {H.}~\bibnamefont {Su}}, \bibinfo
  {author} {\bibfnamefont {K.}~\bibnamefont {Mo}}, \ and\ \bibinfo {author}
  {\bibfnamefont {L.~J.}\ \bibnamefont {Guibas}},\ }\bibfield  {title}
  {\enquote {\bibinfo {title} {Pointnet: Deep learning on point sets for 3d
  classification and segmentation},}\ }in\ \href@noop {} {\emph {\bibinfo
  {booktitle} {Proceedings of the IEEE conference on computer vision and
  pattern recognition}}}\ (\bibinfo {year} {2017})\ pp.\ \bibinfo {pages}
  {652--660}\BibitemShut {NoStop}%
\bibitem [{\citenamefont {Thomas}\ \emph {et~al.}(2019)\citenamefont {Thomas},
  \citenamefont {Qi}, \citenamefont {Deschaud}, \citenamefont {Marcotegui},
  \citenamefont {Goulette},\ and\ \citenamefont {Guibas}}]{thomas2019kpconv}%
  \BibitemOpen
  \bibfield  {author} {\bibinfo {author} {\bibfnamefont {H.}~\bibnamefont
  {Thomas}}, \bibinfo {author} {\bibfnamefont {C.~R.}\ \bibnamefont {Qi}},
  \bibinfo {author} {\bibfnamefont {J.-E.}\ \bibnamefont {Deschaud}}, \bibinfo
  {author} {\bibfnamefont {B.}~\bibnamefont {Marcotegui}}, \bibinfo {author}
  {\bibfnamefont {F.}~\bibnamefont {Goulette}}, \ and\ \bibinfo {author}
  {\bibfnamefont {L.~J.}\ \bibnamefont {Guibas}},\ }\bibfield  {title}
  {\enquote {\bibinfo {title} {Kpconv: Flexible and deformable convolution for
  point clouds},}\ }in\ \href@noop {} {\emph {\bibinfo {booktitle} {Proceedings
  of the IEEE/CVF International Conference on Computer Vision}}}\ (\bibinfo
  {year} {2019})\ pp.\ \bibinfo {pages} {6411--6420}\BibitemShut {NoStop}%
\bibitem [{\citenamefont {Qi}\ \emph {et~al.}(2019)\citenamefont {Qi},
  \citenamefont {Litany}, \citenamefont {He},\ and\ \citenamefont
  {Guibas}}]{qi2019deep}%
  \BibitemOpen
  \bibfield  {author} {\bibinfo {author} {\bibfnamefont {C.~R.}\ \bibnamefont
  {Qi}}, \bibinfo {author} {\bibfnamefont {O.}~\bibnamefont {Litany}}, \bibinfo
  {author} {\bibfnamefont {K.}~\bibnamefont {He}}, \ and\ \bibinfo {author}
  {\bibfnamefont {L.~J.}\ \bibnamefont {Guibas}},\ }\bibfield  {title}
  {\enquote {\bibinfo {title} {Deep hough voting for 3d object detection in
  point clouds},}\ }in\ \href@noop {} {\emph {\bibinfo {booktitle} {Proceedings
  of the IEEE/CVF International Conference on Computer Vision}}}\ (\bibinfo
  {year} {2019})\ pp.\ \bibinfo {pages} {9277--9286}\BibitemShut {NoStop}%
\bibitem [{\citenamefont {Qi}\ \emph {et~al.}(2018)\citenamefont {Qi},
  \citenamefont {Liu}, \citenamefont {Wu}, \citenamefont {Su},\ and\
  \citenamefont {Guibas}}]{qi2018frustum}%
  \BibitemOpen
  \bibfield  {author} {\bibinfo {author} {\bibfnamefont {C.~R.}\ \bibnamefont
  {Qi}}, \bibinfo {author} {\bibfnamefont {W.}~\bibnamefont {Liu}}, \bibinfo
  {author} {\bibfnamefont {C.}~\bibnamefont {Wu}}, \bibinfo {author}
  {\bibfnamefont {H.}~\bibnamefont {Su}}, \ and\ \bibinfo {author}
  {\bibfnamefont {L.~J.}\ \bibnamefont {Guibas}},\ }\bibfield  {title}
  {\enquote {\bibinfo {title} {Frustum pointnets for 3d object detection from
  rgb-d data},}\ }in\ \href@noop {} {\emph {\bibinfo {booktitle} {Proceedings
  of the IEEE conference on computer vision and pattern recognition}}}\
  (\bibinfo {year} {2018})\ pp.\ \bibinfo {pages} {918--927}\BibitemShut
  {NoStop}%
\bibitem [{\citenamefont {Liu}, \citenamefont {Qi},\ and\ \citenamefont
  {Guibas}(2019)}]{liu2019flownet3d}%
  \BibitemOpen
  \bibfield  {author} {\bibinfo {author} {\bibfnamefont {X.}~\bibnamefont
  {Liu}}, \bibinfo {author} {\bibfnamefont {C.~R.}\ \bibnamefont {Qi}}, \ and\
  \bibinfo {author} {\bibfnamefont {L.~J.}\ \bibnamefont {Guibas}},\ }\bibfield
   {title} {\enquote {\bibinfo {title} {Flownet3d: Learning scene flow in 3d
  point clouds},}\ }in\ \href@noop {} {\emph {\bibinfo {booktitle} {Proceedings
  of the IEEE/CVF Conference on Computer Vision and Pattern Recognition}}}\
  (\bibinfo {year} {2019})\ pp.\ \bibinfo {pages} {529--537}\BibitemShut
  {NoStop}%
\bibitem [{\citenamefont {Rempe}\ \emph {et~al.}(2020)\citenamefont {Rempe},
  \citenamefont {Birdal}, \citenamefont {Zhao}, \citenamefont {Gojcic},
  \citenamefont {Sridhar},\ and\ \citenamefont
  {Guibas}}]{DBLP1:journals/corr/abs-2008-02792}%
  \BibitemOpen
  \bibfield  {author} {\bibinfo {author} {\bibfnamefont {D.}~\bibnamefont
  {Rempe}}, \bibinfo {author} {\bibfnamefont {T.}~\bibnamefont {Birdal}},
  \bibinfo {author} {\bibfnamefont {Y.}~\bibnamefont {Zhao}}, \bibinfo {author}
  {\bibfnamefont {Z.}~\bibnamefont {Gojcic}}, \bibinfo {author} {\bibfnamefont
  {S.}~\bibnamefont {Sridhar}}, \ and\ \bibinfo {author} {\bibfnamefont
  {L.~J.}\ \bibnamefont {Guibas}},\ }\bibfield  {title} {\enquote {\bibinfo
  {title} {Caspr: Learning canonical spatiotemporal point cloud
  representations},}\ }\href {https://arxiv.org/abs/2008.02792} {\bibfield
  {journal} {\bibinfo  {journal} {CoRR}\ }\textbf {\bibinfo {volume}
  {abs/2008.02792}} (\bibinfo {year} {2020})},\ \Eprint
  {http://arxiv.org/abs/2008.02792} {arXiv:2008.02792} \BibitemShut {NoStop}%
\bibitem [{\citenamefont {Kashefi}, \citenamefont {Rempe},\ and\ \citenamefont
  {Guibas}(2021)}]{kashefi2021point}%
  \BibitemOpen
  \bibfield  {author} {\bibinfo {author} {\bibfnamefont {A.}~\bibnamefont
  {Kashefi}}, \bibinfo {author} {\bibfnamefont {D.}~\bibnamefont {Rempe}}, \
  and\ \bibinfo {author} {\bibfnamefont {L.~J.}\ \bibnamefont {Guibas}},\
  }\bibfield  {title} {\enquote {\bibinfo {title} {A point-cloud deep learning
  framework for prediction of fluid flow fields on irregular geometries},}\
  }\href@noop {} {\bibfield  {journal} {\bibinfo  {journal} {Physics of
  Fluids}\ }\textbf {\bibinfo {volume} {33}},\ \bibinfo {pages} {027104}
  (\bibinfo {year} {2021})}\BibitemShut {NoStop}%
\bibitem [{\citenamefont {Rempe}\ \emph {et~al.}(2019)\citenamefont {Rempe},
  \citenamefont {Sridhar}, \citenamefont {Wang},\ and\ \citenamefont
  {Guibas}}]{DBLP2:conf/cvpr/Rempe0WG19}%
  \BibitemOpen
  \bibfield  {author} {\bibinfo {author} {\bibfnamefont {D.}~\bibnamefont
  {Rempe}}, \bibinfo {author} {\bibfnamefont {S.}~\bibnamefont {Sridhar}},
  \bibinfo {author} {\bibfnamefont {H.}~\bibnamefont {Wang}}, \ and\ \bibinfo
  {author} {\bibfnamefont {L.~J.}\ \bibnamefont {Guibas}},\ }\bibfield  {title}
  {\enquote {\bibinfo {title} {Learning generalizable final-state dynamics of
  3d rigid objects},}\ }in\ \href
  {http://openaccess.thecvf.com/content\_CVPRW\_2019/html/Vision\_Meets\_Cognition\_Camera\_Ready/Rempe\_Learning\_Generalizable\_Final-State\_Dynamics\_of\_3D\_Rigid\_Objects\_CVPRW\_2019\_paper.html}
  {\emph {\bibinfo {booktitle} {{IEEE} Conference on Computer Vision and
  Pattern Recognition Workshops, {CVPR} Workshops 2019, Long Beach, CA, USA,
  June 16-20, 2019}}}\ (\bibinfo  {publisher} {Computer Vision Foundation /
  {IEEE}},\ \bibinfo {year} {2019})\ pp.\ \bibinfo {pages} {17--20}\BibitemShut
  {NoStop}%
\bibitem [{\citenamefont {DeFever}\ \emph {et~al.}(2019)\citenamefont
  {DeFever}, \citenamefont {Targonski}, \citenamefont {Hall}, \citenamefont
  {Smith},\ and\ \citenamefont {Sarupria}}]{defever2019generalized}%
  \BibitemOpen
  \bibfield  {author} {\bibinfo {author} {\bibfnamefont {R.~S.}\ \bibnamefont
  {DeFever}}, \bibinfo {author} {\bibfnamefont {C.}~\bibnamefont {Targonski}},
  \bibinfo {author} {\bibfnamefont {S.~W.}\ \bibnamefont {Hall}}, \bibinfo
  {author} {\bibfnamefont {M.~C.}\ \bibnamefont {Smith}}, \ and\ \bibinfo
  {author} {\bibfnamefont {S.}~\bibnamefont {Sarupria}},\ }\bibfield  {title}
  {\enquote {\bibinfo {title} {A generalized deep learning approach for local
  structure identification in molecular simulations},}\ }\href@noop {}
  {\bibfield  {journal} {\bibinfo  {journal} {Chemical science}\ }\textbf
  {\bibinfo {volume} {10}},\ \bibinfo {pages} {7503--7515} (\bibinfo {year}
  {2019})}\BibitemShut {NoStop}%
\bibitem [{\citenamefont {Keehm}, \citenamefont {Mukerji},\ and\ \citenamefont
  {Nur}(2004)}]{keehm2004permeability}%
  \BibitemOpen
  \bibfield  {author} {\bibinfo {author} {\bibfnamefont {Y.}~\bibnamefont
  {Keehm}}, \bibinfo {author} {\bibfnamefont {T.}~\bibnamefont {Mukerji}}, \
  and\ \bibinfo {author} {\bibfnamefont {A.}~\bibnamefont {Nur}},\ }\bibfield
  {title} {\enquote {\bibinfo {title} {Permeability prediction from thin
  sections: 3d reconstruction and lattice-boltzmann flow simulation},}\
  }\href@noop {} {\bibfield  {journal} {\bibinfo  {journal} {Geophysical
  Research Letters}\ }\textbf {\bibinfo {volume} {31}} (\bibinfo {year}
  {2004})}\BibitemShut {NoStop}%
\bibitem [{\citenamefont {Darcy}(1856)}]{darcy1856fontaines}%
  \BibitemOpen
  \bibfield  {author} {\bibinfo {author} {\bibfnamefont {H.}~\bibnamefont
  {Darcy}},\ }\href@noop {} {\emph {\bibinfo {title} {Les fontaines publiques
  de la ville de Dijon: exposition et application...}}}\ (\bibinfo  {publisher}
  {Victor Dalmont},\ \bibinfo {year} {1856})\BibitemShut {NoStop}%
\bibitem [{\citenamefont {Eshghinejadfard}\ \emph {et~al.}(2016)\citenamefont
  {Eshghinejadfard}, \citenamefont {Dar{\'o}czy}, \citenamefont {Janiga},\ and\
  \citenamefont {Th{\'e}venin}}]{eshghinejadfard2016calculation}%
  \BibitemOpen
  \bibfield  {author} {\bibinfo {author} {\bibfnamefont {A.}~\bibnamefont
  {Eshghinejadfard}}, \bibinfo {author} {\bibfnamefont {L.}~\bibnamefont
  {Dar{\'o}czy}}, \bibinfo {author} {\bibfnamefont {G.}~\bibnamefont {Janiga}},
  \ and\ \bibinfo {author} {\bibfnamefont {D.}~\bibnamefont {Th{\'e}venin}},\
  }\bibfield  {title} {\enquote {\bibinfo {title} {Calculation of the
  permeability in porous media using the lattice boltzmann method},}\
  }\href@noop {} {\bibfield  {journal} {\bibinfo  {journal} {International
  Journal of Heat and Fluid Flow}\ }\textbf {\bibinfo {volume} {62}},\ \bibinfo
  {pages} {93--103} (\bibinfo {year} {2016})}\BibitemShut {NoStop}%
\bibitem [{\citenamefont
  {Lantu{\'e}joul}(2013)}]{lantuejoul2013geostatistical}%
  \BibitemOpen
  \bibfield  {author} {\bibinfo {author} {\bibfnamefont {C.}~\bibnamefont
  {Lantu{\'e}joul}},\ }\href@noop {} {\emph {\bibinfo {title} {Geostatistical
  simulation: models and algorithms}}}\ (\bibinfo  {publisher} {Springer
  Science \& Business Media},\ \bibinfo {year} {2013})\BibitemShut {NoStop}%
\bibitem [{\citenamefont {Xu}, \citenamefont {Journel}\ \emph
  {et~al.}(1993)\citenamefont {Xu}, \citenamefont {Journel} \emph
  {et~al.}}]{xu1993gtsim}%
  \BibitemOpen
  \bibfield  {author} {\bibinfo {author} {\bibfnamefont {W.}~\bibnamefont
  {Xu}}, \bibinfo {author} {\bibfnamefont {A.}~\bibnamefont {Journel}},  \emph
  {et~al.},\ }\bibfield  {title} {\enquote {\bibinfo {title} {Gtsim: Gaussian
  truncated simulations of reservoir units in a w. texas carbonate field},}\
  }\href@noop {} {\bibfield  {journal} {\bibinfo  {journal} {paper SPE}\
  }\textbf {\bibinfo {volume} {27412}},\ \bibinfo {pages} {3--6} (\bibinfo
  {year} {1993})}\BibitemShut {NoStop}%
\bibitem [{\citenamefont {Ioffe}\ and\ \citenamefont
  {Szegedy}(2015)}]{ioffe2015batch}%
  \BibitemOpen
  \bibfield  {author} {\bibinfo {author} {\bibfnamefont {S.}~\bibnamefont
  {Ioffe}}\ and\ \bibinfo {author} {\bibfnamefont {C.}~\bibnamefont
  {Szegedy}},\ }\bibfield  {title} {\enquote {\bibinfo {title} {Batch
  normalization: Accelerating deep network training by reducing internal
  covariate shift},}\ }in\ \href@noop {} {\emph {\bibinfo {booktitle}
  {International conference on machine learning}}}\ (\bibinfo {organization}
  {PMLR},\ \bibinfo {year} {2015})\ pp.\ \bibinfo {pages}
  {448--456}\BibitemShut {NoStop}%
\bibitem [{\citenamefont {Yu}\ \emph {et~al.}(2017)\citenamefont {Yu},
  \citenamefont {Gong}, \citenamefont {Zhong},\ and\ \citenamefont
  {Shan}}]{yu2017unsupervised}%
  \BibitemOpen
  \bibfield  {author} {\bibinfo {author} {\bibfnamefont {Y.}~\bibnamefont
  {Yu}}, \bibinfo {author} {\bibfnamefont {Z.}~\bibnamefont {Gong}}, \bibinfo
  {author} {\bibfnamefont {P.}~\bibnamefont {Zhong}}, \ and\ \bibinfo {author}
  {\bibfnamefont {J.}~\bibnamefont {Shan}},\ }\bibfield  {title} {\enquote
  {\bibinfo {title} {Unsupervised representation learning with deep
  convolutional neural network for remote sensing images},}\ }in\ \href@noop {}
  {\emph {\bibinfo {booktitle} {International Conference on Image and
  Graphics}}}\ (\bibinfo {organization} {Springer},\ \bibinfo {year} {2017})\
  pp.\ \bibinfo {pages} {97--108}\BibitemShut {NoStop}%
\bibitem [{\citenamefont {Ajit}, \citenamefont {Acharya},\ and\ \citenamefont
  {Samanta}(2020)}]{ajit2020review}%
  \BibitemOpen
  \bibfield  {author} {\bibinfo {author} {\bibfnamefont {A.}~\bibnamefont
  {Ajit}}, \bibinfo {author} {\bibfnamefont {K.}~\bibnamefont {Acharya}}, \
  and\ \bibinfo {author} {\bibfnamefont {A.}~\bibnamefont {Samanta}},\
  }\bibfield  {title} {\enquote {\bibinfo {title} {A review of convolutional
  neural networks},}\ }in\ \href@noop {} {\emph {\bibinfo {booktitle} {2020
  International Conference on Emerging Trends in Information Technology and
  Engineering (ic-ETITE)}}}\ (\bibinfo {organization} {IEEE},\ \bibinfo {year}
  {2020})\ pp.\ \bibinfo {pages} {1--5}\BibitemShut {NoStop}%
\bibitem [{\citenamefont {Khan}\ \emph {et~al.}(2020)\citenamefont {Khan},
  \citenamefont {Sohail}, \citenamefont {Zahoora},\ and\ \citenamefont
  {Qureshi}}]{khan2020survey}%
  \BibitemOpen
  \bibfield  {author} {\bibinfo {author} {\bibfnamefont {A.}~\bibnamefont
  {Khan}}, \bibinfo {author} {\bibfnamefont {A.}~\bibnamefont {Sohail}},
  \bibinfo {author} {\bibfnamefont {U.}~\bibnamefont {Zahoora}}, \ and\
  \bibinfo {author} {\bibfnamefont {A.~S.}\ \bibnamefont {Qureshi}},\
  }\bibfield  {title} {\enquote {\bibinfo {title} {A survey of the recent
  architectures of deep convolutional neural networks},}\ }\href@noop {}
  {\bibfield  {journal} {\bibinfo  {journal} {Artificial Intelligence Review}\
  }\textbf {\bibinfo {volume} {53}},\ \bibinfo {pages} {5455--5516} (\bibinfo
  {year} {2020})}\BibitemShut {NoStop}%
\bibitem [{\citenamefont {Nagi}\ \emph {et~al.}(2011)\citenamefont {Nagi},
  \citenamefont {Ducatelle}, \citenamefont {Di~Caro}, \citenamefont
  {Cire{\c{s}}an}, \citenamefont {Meier}, \citenamefont {Giusti}, \citenamefont
  {Nagi}, \citenamefont {Schmidhuber},\ and\ \citenamefont
  {Gambardella}}]{nagi2011max}%
  \BibitemOpen
  \bibfield  {author} {\bibinfo {author} {\bibfnamefont {J.}~\bibnamefont
  {Nagi}}, \bibinfo {author} {\bibfnamefont {F.}~\bibnamefont {Ducatelle}},
  \bibinfo {author} {\bibfnamefont {G.~A.}\ \bibnamefont {Di~Caro}}, \bibinfo
  {author} {\bibfnamefont {D.}~\bibnamefont {Cire{\c{s}}an}}, \bibinfo {author}
  {\bibfnamefont {U.}~\bibnamefont {Meier}}, \bibinfo {author} {\bibfnamefont
  {A.}~\bibnamefont {Giusti}}, \bibinfo {author} {\bibfnamefont
  {F.}~\bibnamefont {Nagi}}, \bibinfo {author} {\bibfnamefont {J.}~\bibnamefont
  {Schmidhuber}}, \ and\ \bibinfo {author} {\bibfnamefont {L.~M.}\ \bibnamefont
  {Gambardella}},\ }\bibfield  {title} {\enquote {\bibinfo {title} {Max-pooling
  convolutional neural networks for vision-based hand gesture recognition},}\
  }in\ \href@noop {} {\emph {\bibinfo {booktitle} {2011 IEEE International
  Conference on Signal and Image Processing Applications (ICSIPA)}}}\ (\bibinfo
  {organization} {IEEE},\ \bibinfo {year} {2011})\ pp.\ \bibinfo {pages}
  {342--347}\BibitemShut {NoStop}%
\bibitem [{\citenamefont {Wang}\ \emph {et~al.}(2018)\citenamefont {Wang},
  \citenamefont {Chen}, \citenamefont {Yuan}, \citenamefont {Liu},
  \citenamefont {Huang}, \citenamefont {Hou},\ and\ \citenamefont
  {Cottrell}}]{wang2018understanding}%
  \BibitemOpen
  \bibfield  {author} {\bibinfo {author} {\bibfnamefont {P.}~\bibnamefont
  {Wang}}, \bibinfo {author} {\bibfnamefont {P.}~\bibnamefont {Chen}}, \bibinfo
  {author} {\bibfnamefont {Y.}~\bibnamefont {Yuan}}, \bibinfo {author}
  {\bibfnamefont {D.}~\bibnamefont {Liu}}, \bibinfo {author} {\bibfnamefont
  {Z.}~\bibnamefont {Huang}}, \bibinfo {author} {\bibfnamefont
  {X.}~\bibnamefont {Hou}}, \ and\ \bibinfo {author} {\bibfnamefont
  {G.}~\bibnamefont {Cottrell}},\ }\bibfield  {title} {\enquote {\bibinfo
  {title} {Understanding convolution for semantic segmentation},}\ }in\
  \href@noop {} {\emph {\bibinfo {booktitle} {2018 IEEE winter conference on
  applications of computer vision (WACV)}}}\ (\bibinfo {organization} {IEEE},\
  \bibinfo {year} {2018})\ pp.\ \bibinfo {pages} {1451--1460}\BibitemShut
  {NoStop}%
\bibitem [{\citenamefont {Yamanaka}, \citenamefont {Kuwashima},\ and\
  \citenamefont {Kurita}(2017)}]{yamanaka2017fast}%
  \BibitemOpen
  \bibfield  {author} {\bibinfo {author} {\bibfnamefont {J.}~\bibnamefont
  {Yamanaka}}, \bibinfo {author} {\bibfnamefont {S.}~\bibnamefont {Kuwashima}},
  \ and\ \bibinfo {author} {\bibfnamefont {T.}~\bibnamefont {Kurita}},\
  }\bibfield  {title} {\enquote {\bibinfo {title} {Fast and accurate image
  super resolution by deep cnn with skip connection and network in network},}\
  }in\ \href@noop {} {\emph {\bibinfo {booktitle} {International Conference on
  Neural Information Processing}}}\ (\bibinfo {organization} {Springer},\
  \bibinfo {year} {2017})\ pp.\ \bibinfo {pages} {217--225}\BibitemShut
  {NoStop}%
\bibitem [{\citenamefont {Sekar}\ \emph {et~al.}(2019)\citenamefont {Sekar},
  \citenamefont {Jiang}, \citenamefont {Shu},\ and\ \citenamefont
  {Khoo}}]{sekar2019fast}%
  \BibitemOpen
  \bibfield  {author} {\bibinfo {author} {\bibfnamefont {V.}~\bibnamefont
  {Sekar}}, \bibinfo {author} {\bibfnamefont {Q.}~\bibnamefont {Jiang}},
  \bibinfo {author} {\bibfnamefont {C.}~\bibnamefont {Shu}}, \ and\ \bibinfo
  {author} {\bibfnamefont {B.~C.}\ \bibnamefont {Khoo}},\ }\bibfield  {title}
  {\enquote {\bibinfo {title} {Fast flow field prediction over airfoils using
  deep learning approach},}\ }\href@noop {} {\bibfield  {journal} {\bibinfo
  {journal} {Physics of Fluids}\ }\textbf {\bibinfo {volume} {31}},\ \bibinfo
  {pages} {057103} (\bibinfo {year} {2019})}\BibitemShut {NoStop}%
\bibitem [{\citenamefont {Kingma}\ and\ \citenamefont
  {Ba}(2014)}]{kingma2014adam}%
  \BibitemOpen
  \bibfield  {author} {\bibinfo {author} {\bibfnamefont {D.~P.}\ \bibnamefont
  {Kingma}}\ and\ \bibinfo {author} {\bibfnamefont {J.}~\bibnamefont {Ba}},\
  }\bibfield  {title} {\enquote {\bibinfo {title} {Adam: A method for
  stochastic optimization},}\ }\href@noop {} {\bibfield  {journal} {\bibinfo
  {journal} {arXiv preprint arXiv:1412.6980}\ } (\bibinfo {year}
  {2014})}\BibitemShut {NoStop}%
\bibitem [{\citenamefont {Jain}\ and\ \citenamefont {Kar}(2017)}]{jain2017non}%
  \BibitemOpen
  \bibfield  {author} {\bibinfo {author} {\bibfnamefont {P.}~\bibnamefont
  {Jain}}\ and\ \bibinfo {author} {\bibfnamefont {P.}~\bibnamefont {Kar}},\
  }\bibfield  {title} {\enquote {\bibinfo {title} {Non-convex optimization for
  machine learning},}\ }\href@noop {} {\bibfield  {journal} {\bibinfo
  {journal} {arXiv preprint arXiv:1712.07897}\ } (\bibinfo {year}
  {2017})}\BibitemShut {NoStop}%
\bibitem [{\citenamefont {Ronneberger}, \citenamefont {Fischer},\ and\
  \citenamefont {Brox}(2015)}]{ronneberger2015u}%
  \BibitemOpen
  \bibfield  {author} {\bibinfo {author} {\bibfnamefont {O.}~\bibnamefont
  {Ronneberger}}, \bibinfo {author} {\bibfnamefont {P.}~\bibnamefont
  {Fischer}}, \ and\ \bibinfo {author} {\bibfnamefont {T.}~\bibnamefont
  {Brox}},\ }\bibfield  {title} {\enquote {\bibinfo {title} {U-net:
  Convolutional networks for biomedical image segmentation},}\ }in\ \href@noop
  {} {\emph {\bibinfo {booktitle} {International Conference on Medical image
  computing and computer-assisted intervention}}}\ (\bibinfo {organization}
  {Springer},\ \bibinfo {year} {2015})\ pp.\ \bibinfo {pages}
  {234--241}\BibitemShut {NoStop}%
\bibitem [{\citenamefont {Bhatnagar}\ \emph {et~al.}(2019)\citenamefont
  {Bhatnagar}, \citenamefont {Afshar}, \citenamefont {Pan}, \citenamefont
  {Duraisamy},\ and\ \citenamefont {Kaushik}}]{bhatnagar2019prediction}%
  \BibitemOpen
  \bibfield  {author} {\bibinfo {author} {\bibfnamefont {S.}~\bibnamefont
  {Bhatnagar}}, \bibinfo {author} {\bibfnamefont {Y.}~\bibnamefont {Afshar}},
  \bibinfo {author} {\bibfnamefont {S.}~\bibnamefont {Pan}}, \bibinfo {author}
  {\bibfnamefont {K.}~\bibnamefont {Duraisamy}}, \ and\ \bibinfo {author}
  {\bibfnamefont {S.}~\bibnamefont {Kaushik}},\ }\bibfield  {title} {\enquote
  {\bibinfo {title} {Prediction of aerodynamic flow fields using convolutional
  neural networks},}\ }\href@noop {} {\bibfield  {journal} {\bibinfo  {journal}
  {Computational Mechanics}\ }\textbf {\bibinfo {volume} {64}},\ \bibinfo
  {pages} {525--545} (\bibinfo {year} {2019})}\BibitemShut {NoStop}%
\bibitem [{\citenamefont {Qi}\ \emph {et~al.}(2017{\natexlab{b}})\citenamefont
  {Qi}, \citenamefont {Yi}, \citenamefont {Su},\ and\ \citenamefont
  {Guibas}}]{P++}%
  \BibitemOpen
  \bibfield  {author} {\bibinfo {author} {\bibfnamefont {C.~R.}\ \bibnamefont
  {Qi}}, \bibinfo {author} {\bibfnamefont {L.}~\bibnamefont {Yi}}, \bibinfo
  {author} {\bibfnamefont {H.}~\bibnamefont {Su}}, \ and\ \bibinfo {author}
  {\bibfnamefont {L.~J.}\ \bibnamefont {Guibas}},\ }\bibfield  {title}
  {\enquote {\bibinfo {title} {Pointnet++: Deep hierarchical feature learning
  on point sets in a metric space},}\ }in\ \href
  {https://proceedings.neurips.cc/paper/2017/hash/d8bf84be3800d12f74d8b05e9b89836f-Abstract.html}
  {\emph {\bibinfo {booktitle} {Advances in Neural Information Processing
  Systems 30: Annual Conference on Neural Information Processing Systems 2017,
  December 4-9, 2017, Long Beach, CA, {USA}}}},\ \bibinfo {editor} {edited by\
  \bibinfo {editor} {\bibfnamefont {I.}~\bibnamefont {Guyon}}, \bibinfo
  {editor} {\bibfnamefont {U.}~\bibnamefont {von Luxburg}}, \bibinfo {editor}
  {\bibfnamefont {S.}~\bibnamefont {Bengio}}, \bibinfo {editor} {\bibfnamefont
  {H.~M.}\ \bibnamefont {Wallach}}, \bibinfo {editor} {\bibfnamefont
  {R.}~\bibnamefont {Fergus}}, \bibinfo {editor} {\bibfnamefont {S.~V.~N.}\
  \bibnamefont {Vishwanathan}}, \ and\ \bibinfo {editor} {\bibfnamefont
  {R.}~\bibnamefont {Garnett}}}\ (\bibinfo {year} {2017})\ pp.\ \bibinfo
  {pages} {5099--5108}\BibitemShut {NoStop}%
\bibitem [{\citenamefont {Subramaniam}\ \emph {et~al.}(2020)\citenamefont
  {Subramaniam}, \citenamefont {Wong}, \citenamefont {Borker}, \citenamefont
  {Nimmagadda},\ and\ \citenamefont {Lele}}]{subramaniam2020turbulence}%
  \BibitemOpen
  \bibfield  {author} {\bibinfo {author} {\bibfnamefont {A.}~\bibnamefont
  {Subramaniam}}, \bibinfo {author} {\bibfnamefont {M.~L.}\ \bibnamefont
  {Wong}}, \bibinfo {author} {\bibfnamefont {R.~D.}\ \bibnamefont {Borker}},
  \bibinfo {author} {\bibfnamefont {S.}~\bibnamefont {Nimmagadda}}, \ and\
  \bibinfo {author} {\bibfnamefont {S.~K.}\ \bibnamefont {Lele}},\ }\bibfield
  {title} {\enquote {\bibinfo {title} {Turbulence enrichment using
  physics-informed generative adversarial networks},}\ }\href@noop {}
  {\bibfield  {journal} {\bibinfo  {journal} {arXiv preprint arXiv:2003.01907}\
  } (\bibinfo {year} {2020})}\BibitemShut {NoStop}%
\bibitem [{\citenamefont {Jagtap}, \citenamefont {Kharazmi},\ and\
  \citenamefont {Karniadakis}(2020)}]{jagtap2020conservative}%
  \BibitemOpen
  \bibfield  {author} {\bibinfo {author} {\bibfnamefont {A.~D.}\ \bibnamefont
  {Jagtap}}, \bibinfo {author} {\bibfnamefont {E.}~\bibnamefont {Kharazmi}}, \
  and\ \bibinfo {author} {\bibfnamefont {G.~E.}\ \bibnamefont {Karniadakis}},\
  }\bibfield  {title} {\enquote {\bibinfo {title} {Conservative
  physics-informed neural networks on discrete domains for conservation laws:
  Applications to forward and inverse problems},}\ }\href@noop {} {\bibfield
  {journal} {\bibinfo  {journal} {Computer Methods in Applied Mechanics and
  Engineering}\ }\textbf {\bibinfo {volume} {365}},\ \bibinfo {pages} {113028}
  (\bibinfo {year} {2020})}\BibitemShut {NoStop}%
\bibitem [{\citenamefont {Mao}, \citenamefont {Jagtap},\ and\ \citenamefont
  {Karniadakis}(2020)}]{mao2020physics}%
  \BibitemOpen
  \bibfield  {author} {\bibinfo {author} {\bibfnamefont {Z.}~\bibnamefont
  {Mao}}, \bibinfo {author} {\bibfnamefont {A.~D.}\ \bibnamefont {Jagtap}}, \
  and\ \bibinfo {author} {\bibfnamefont {G.~E.}\ \bibnamefont {Karniadakis}},\
  }\bibfield  {title} {\enquote {\bibinfo {title} {Physics-informed neural
  networks for high-speed flows},}\ }\href@noop {} {\bibfield  {journal}
  {\bibinfo  {journal} {Computer Methods in Applied Mechanics and Engineering}\
  }\textbf {\bibinfo {volume} {360}},\ \bibinfo {pages} {112789} (\bibinfo
  {year} {2020})}\BibitemShut {NoStop}%
\bibitem [{\citenamefont {Raissi}, \citenamefont {Perdikaris},\ and\
  \citenamefont {Karniadakis}(2019)}]{raissi2019physics}%
  \BibitemOpen
  \bibfield  {author} {\bibinfo {author} {\bibfnamefont {M.}~\bibnamefont
  {Raissi}}, \bibinfo {author} {\bibfnamefont {P.}~\bibnamefont {Perdikaris}},
  \ and\ \bibinfo {author} {\bibfnamefont {G.~E.}\ \bibnamefont
  {Karniadakis}},\ }\bibfield  {title} {\enquote {\bibinfo {title}
  {Physics-informed neural networks: A deep learning framework for solving
  forward and inverse problems involving nonlinear partial differential
  equations},}\ }\href@noop {} {\bibfield  {journal} {\bibinfo  {journal}
  {Journal of Computational Physics}\ }\textbf {\bibinfo {volume} {378}},\
  \bibinfo {pages} {686--707} (\bibinfo {year} {2019})}\BibitemShut {NoStop}%
\end{thebibliography}%

\end{document}